\documentclass[12pt,preprint]{aastex}

\begin{document}

\title{Relativistic Jet Properties of GeV-TeV Blazars and Possible Implications for the Jet Formation, Composition, and Cavity Kinematics}
\author{Jin Zhang\altaffilmark{1,3}, Xiao-Na Sun\altaffilmark{2}, En-Wei Liang\altaffilmark{2,1,3}, Rui-Jing Lu\altaffilmark{2}, Ye Lu\altaffilmark{1}, Shuang-Nan Zhang\altaffilmark{1,4,5}}
\altaffiltext{1}{National Astronomical Observatories, Chinese Academy of Sciences, Beijing 100012, China; zhang.jin@hotmail.com}\altaffiltext{2}{Department of Physics and GXU-NAOC Center for Astrophysics
and Space Sciences, Guangxi University, Nanning 530004, China; lew@gxu.edu.cn} \altaffiltext{3}{Key Laboratory for the Structure and Evolution of Celestial Objects, Chinese Academy of Sciences, Kunming, 650011, China}\altaffiltext{4}{Key Laboratory of Particle
Astrophysics, Institute of High Energy Physics, Chinese Academy of Sciences, Beijing 100049,
China}\altaffiltext{5}{Physics Department, University of Alabama in
Huntsville, Huntsville, AL 35899, USA}
\begin{abstract}
We fit the spectral energy distributions (SEDs) of a GeV-TeV FSRQ sample with the leptonic model. Their $\gamma_{\min}$ of the relativistic electron distributions, which significantly affect the estimates of the jet properties, are constrained, with a typical value of $\sim$ 48. Their jet power, magnetized parameter, radiation efficiency, and jet production/radiation rates per central black hole (BH) mass are derived and compared to that of BL Lacs. We show that the FSRQ jets may be dominated by the Poynting flux and have a high radiation efficiency, whereas the BL Lac jets are likely dominated by particles and have a lower radiation efficiency than FSRQs. Being different from BL Lacs, the jet powers of FSRQs are proportional to their central BH masses. The jet production and radiation rates of the FSRQs distribute in narrow ranges and are correlated with each other, whereas no similar feature is found for the BL Lacs. We also show that the jet power is correlated with the cavity kinetic power, and the magnetic field energy in the jets may provide the cavity kinetic energy of FSRQs and the kinetic energy of cold protons in the jets may be crucial for cavity kinetic energy of BL Lacs. We suggest that the dominating formation mechanism of FSRQ jets may be the BZ process, but BL Lac jets may be produced via the BP and/or BZ processes, depending on the structures and accretion rates of accretion disks.

\end{abstract}


\section{Introduction}           
\label{sect:intro}
Most confirmed extragalactic GeV-TeV emitters are blazars, a sub-sample of radio-loud active galactic nuclei (AGNs). They are the main targets for studying the cosmic $\gamma$-ray horizon (e.g., Dominguez et al. 2013), the intergalactic magnetic field (e.g., Dai et al. 2002), the contributions to the extragalactic background radiation (Giommi et al. 2006), etc. Their observed emission is dominated by a relativistic jet, which is generally believed to be beamed toward to the Earth. Their broad band spectral energy distributions (SEDs) from the radio to gamma-ray band are usually bimodal, which can be well explained with the synchrotron radiation and the inverse Compton (IC) scattering of relativistic electrons in the jets (Maraschi et al. 1992; Ghisellini et al. 1996; Sikora et al. 2009; Zhang et al. 2012; Gao et al. 2013; Cao \& Wang 2013a).

Blazars are divided into BL Lac objects (BL Lacs) and flat spectrum radio quasars (FSRQs) according to their emission line features. Those blazars with an equivalent width (EW) of the emission lines in the rest frame being narrower than 5 ${\rm \AA}$ are classified as BL Lacs (e.g., Urry \& Padovani 1995). However, some sources are classified as BL Lacs, but show some properties similar to FSRQs (Sbarufatti et al. 2005; Raiteri et al. 2007; Ghisellini et al. 2011; Giommi et al. 2012). The intrinsically weak broad lines in some sources may be overwhelmed by the beamed non-thermal continuum. Therefore, the EW of emission line alone is not a good indicator to distinguish BL Lacs and FSRQs. It was shown that BL Lacs and FSRQs are well separated into two populations in the $\alpha_{\gamma}-L_{\gamma}$ plane, where $\alpha_{\gamma}$ and $L_{\gamma}$ are the spectral index and luminosity in the gamma-ray band (Ghisellini et al. 2009a; Ackermann et al. 2011). The gamma-ray luminosity may be a proxy for the observed bolometric luminosity to distinguish the two kinds of blazars (Ghisellini et al. 2009a). Note that the contributions of the external Compton (EC) scattering by the photons from broad line regions (BLRs) and/or torus (Sikora et al. 1994; Sikora et al. 2009) usually dominate the emission of the IC peak of FSRQs, being different from BL Lacs. It was also proposed that the existence of a BLR in a blazar may associate with the accretion rate (Nicastro et al. 2003; Ho 2008; Elitzur \& Ho 2009; ). Therefore, the intrinsic differences of the jets between FSRQs and BL Lacs may be related to their central black holes (BHs), accretion disks, and the mechanisms of jet formation (e.g., Giommi et al. 2013).

The formation of a relativistic jet from an accreting BH system is still not well understood. It has long been speculated that the
physics in different BH jet systems may be essentially the same (Mirabel 2004; Zhang 2007). It is generally believed that jet launching is connected with the central BH, accretion disk and corona in a source (Armitage \& Natarajan 1999; Merloni \& Fabian 2002; Cao 2004; Wang et al. 2004; Chen et al. 2012) via either the Blandford-Payne (BP, Blandford \& Payne 1982) and/or Blandford-Znajek (BZ, Blandford \& Znajek 1977) mechanisms. The BP mechanism may power a jet by releasing the gravitational energy of accreting matter which moves toward the BH. The rotational energy of a rapidly rotating BH is essential for the BZ process. Jets may be made of several components, but the basic makeup of jets is still largely unknown. Most models of jet formation predict a Poynting flux dominating energy transport. However, the observations of polarized radio and optical emission indicate that at least some of jet should be made of relativistic electrons which radiate through synchrotron emission (Harris et al. 2006).

Broadband SEDs are critical to reveal the radiation mechanisms and the properties of the radiation regions. Powerful blazars are usually high energy photon emitters. The IC bumps of the observed SEDs of blazars typically peak at the GeV-TeV gamma-ray band. The Large Area Telescope (LAT) on board \emph{Fermi} satellite, which covers an energy band from 20 MeV to $\sim 300$ GeV, presents an unprecedented opportunity to explore the natures of blazars (e.g., Abdo et al. 2010a). We systematically model the observed SEDs for {\em Fermi}/LAT blazars to investigate their jet properties. We derive the physical parameters of the jets with the single zone leptonic model using the minimization $\chi^2$ technique. Our results for {\em Fermi}/LAT GeV-TeV BL Lacs are reported in Zhang et al. (2012). It can be found that with the observed broadband SEDs, the model parameters of some BL Lacs are well constrained (see also Cerruti et al. 2013). In this paper, we continue our analysis to the {\em Fermi}/LAT FSRQs and present a comparison of the jet properties between FSRQs and BL Lacs. We also investigate the jet composition and jet launching mechanism of the two kinds of blazars, as well as the kinetic properties of the cavity.

Our sample and the observed SEDs are presented in Section 2. The model and results of SED fitting are described in Section 3. Comparison of jet power and cavity kinetic power is given in Section 4. Jet composition and radiation efficiency are presented in Section 5. Jet bulk motion and electron spectrum are available in Section 6. The connection between the jet power and the central BH mass is shown in Section 7. Discussion on the jet production, composition, and the cavity kinematics is given in Section 8. A summary of our results is presented in Section 9.

\section{Sample and Data}
\label{sect:data}
\begin{figure*}
\includegraphics[angle=0,scale=0.8]{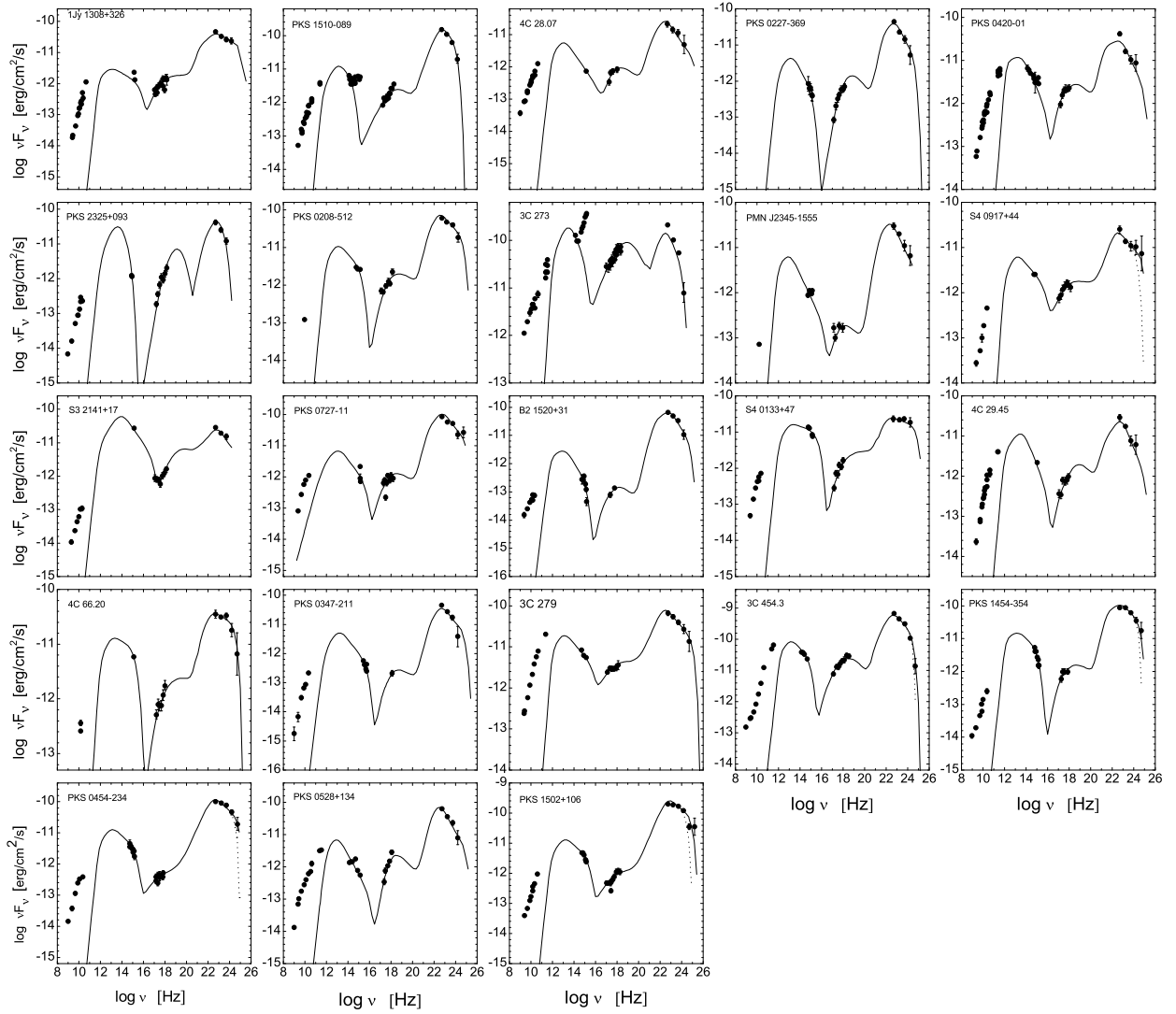}
\caption{Observed SEDs ({\em scattered data points}) with our best model fits ({\em solid lines}) for the FSRQs in our sample. The highest energy points in the SEDs of five sources are only marginally fitted because of the KN effect ({\em dashed lines}). The radio data are not considered in our fits because of the synchrotron self-absorption effect. The blue bumps observed in 3C273 and PKS 1510-089 are also not included since they may be the emission from accretion disk.} \label{SED}
\end{figure*}

{\em Fermi}/LAT bright blazars are selected for our analysis. The BL Lac sample data and the results of their SED fits are available in Zhang et al. (2012).  All the FSRQs in Abdo et al. (2010a) which have confirmed redshift are included in our sample and there are 23 sources. The observed SEDs of these FSRQs are shown in Figure \ref{SED}. Note that more than one observational campaigns with \emph{Swift} were made during the LAT observation period for some sources (3C 279, PKS 0208-512, PKS 0528+134, and 3C 454.3). We thus take the average flux of the data from these campaigns for the SEDs of these sources.

\section{Models and Results of SED Fits for the FSRQs}
The single-zone syn+SSC+EC model is used to fit the SEDs of FSRQs in our sample. Although some groups proposed that the radiation regions of some objects may be outside of the BLR in some states (e.g., Tavecchio et al. 2013), some works for an individual source also indicate that the gamma-ray emitting region is inside the BLR (for 3C 279, Bai et al. 2009; for 3C 454.3, Isler et al. 2013). Since the energy density of torus photon field is much lower than that of BLR, which is about at $\sim10^{-4}$ erg cm$^{-3}$ (Ghisellini et al. 2008) and the EC process of the BLRs may dominate the IC peaks of the SEDs (e.g., Sikora et al. 1994; Ghisellini et al. 2009b; 2010; Chen et al. 2012), we consider only the EC/BLR process. The Klein-Nashina (KN) effects and the absorption of high energy gamma-ray photons by extragalactic background light (Franceschini et al. 2008) are also taken into account in our model calculations. The KN effects are not significant for most of the sources, as shown in Figure \ref{SED}, but the highest energy points in the SEDs of five sources are only marginally fitted with our model\footnote{For the emission above several tens of GeV, it may be contributed by the IC/IR process from a rapidly moving compact blob, and then the two-zone model may be needed to explain this high energy emission as reported by Tavecchio et al. (2011).}. Although a more complex model with more parameters may present better fit to the SEDs, we prefer to use a simple model to fit the SEDs and make statistical analysis for the jet properties based on fitting results.

In order to calculate the energy density of BLR photon field ($U_{\rm BLR}$) in the comoving frame, one needs to estimate the radius ($R_{\rm BLR}$) and total luminosity of the BLR. The radius is usually estimated with the continuum luminosity at $5100{\rm\AA}$ ($L_{5100}$) or the luminosities of emission lines of $\rm H\beta$ and $\rm H\alpha$ ($L_{\rm H\beta}$ and $L_{\rm H\alpha}$), such as $R_{\rm BLR}=(22.3\pm2.1)({L_{5100}}/{10^{44}\rm erg~ s^{-1}})^{0.69\pm0.05}$ (Kaspi et al. 2005), $R_{\rm BLR}=(30.2\pm1.4)({L_{5100}}/{10^{44}\rm erg~s^{-1}})^{0.64\pm0.02}$ (Greene \& Ho 2005), $\log R_{\rm BLR}=-21.3^{+2.9}_{-2.8}+0.519^{+0.063}_{-0.066}\log L_{5100}/{10^{44}\rm erg~ s^{-1}}$ (Bentz et al. 2009), $\log R_{\rm BLR}=(2.22\pm0.02)+(0.51\pm0.02)\log ({L_{\rm H\alpha}}/{10^{44}\rm erg~s^{-1}})$ (Wang \& Zhang 2003), and $\log R_{\rm BLR}=(1.381\pm0.080)+(0.684\pm0.106)\log ({L_{\rm H\beta}}/{10^{42}\rm erg~s^{-1}})$ (Wu et al. 2004), where $R_{\rm BLR}$ is in unit of light days.
$R_{\rm BLR}$ of eight FSRQs in our sample for which the data of emission line are available in literature are estimated with these correlations. The results are listed in Table 1. The estimated $U_{\rm BLR}$ values among these sources vary with a factor of 2 to 3, with an average of $2.07\times10^{-2}$ erg cm$^{-3}$. Note that Ghisellini \& Tavecchio (2008) suggested that $R_{\rm BLR}\sim L_{\rm disk}^{1/2}$ and $L_{\rm BLR}=0.1L_{\rm disk}$, and obtained $U_{\rm BLR}=2.65\times10^{-2}$ erg cm$^{-3}$ at rest frame, which is close to the mean of $U_{\rm BLR}$ in Table 1. Since the BLR is not uniformly and has an angle to the disk, a corrected factor of 17/12 for $U_{\rm BLR}$ is also proposed (Ghisellini \& Madau 1996). The observed energy density in the jet is boosted by a factor of $\Gamma^2$, where $\Gamma$ is the bulk Lorentz factor of the radiating region. Therefore, the estimated $U_{\rm BLR}^{'}$ is about $3.76\times10^{-2}\Gamma^2$ erg cm$^{-3}$ in the comoving frame (Ghisellini et al. 2008; 2009b). In our calculations, we also take $U_{\rm BLR}^{'}=3.76\times10^{-2}\Gamma^2$ erg cm$^{-3}$ for all FSRQs in our sample. We assume $\Gamma=\delta$ since the relativistic jets of blazars are close to the line of sight, where $\delta$ is the beaming factor.

The radiation region is assumed to be a sphere with radius $R$, which is obtained with $R=\delta c\Delta t/(1+z)$, where $\Delta t$ is the variability timescales. Blazars usually have flux variations with timescales from one year to several hours, and even shorter than one hour sometimes (Kartaltepe \& Balonek 2007; Fossati et al. 2008). The timescales for a source at different energy bands may be different (such as for 3C 279, Collmar et al. 2010, Abdo et al. 2010b; for 3C 454.3, Giommi et al. 2006). It was suggested that the rapid variations on timescales shorter than one hour are from an emission blob smaller than the jet cross-section, which moves much faster than the surrounding relativistic jet material (Zacharias \& Schlickeiser 2013). It implies that the rapid flares should originate from local disturbance. Therefore, we do not adopt the observed minimum timescales to estimate the radiating region size. Considering the great uncertainty of the variation timescales and the statistical analysis purpose, we take $\Delta t=12$ hr for all sources, which is the median of the previous BL Lac sample (Zhang et al. 2012).

The electron distribution is taken as a broken power law as used in the community (Ghisellini et al. 2009b; 2010; Chen et al. 2012; Aharonian et al. 2009), which is characterized by an electron density parameter ($N_0$), a break energy $\gamma_{\rm b}$ and indices ($p_1$ and $p_2$) in the range of  $\gamma_{\rm e}\sim[\gamma_{\min}, \gamma_{\max}]$. In case of $\gamma_{\rm e}=1/\sqrt{1-v^{2}/c^{2}}=1$, where $v$ is the velocity of electron, the electron is at rest. Therefore, $\gamma_{\min}$ should be $\geqslant 1$. $\gamma_{\max}$ is usually poorly constrained, but it does not significantly affect our results and is fixed at a large value. Note that $\gamma_{\rm b}$ is not exactly the same as the cooling break since $\gamma_{\rm b}$ results from a complex physical process, including the adiabatic losses, the particle escape, and the cooling (Ghisellini et al. 2009b). We do not concern the physical reasons that make the break in the electron spectrum. $p_1$ and $p_2$ are derived from the spectral indices of the observed SEDs as reported by Zhang et al. (2012). $N_{0}$ and $\gamma_{\rm b}$ depend on the synchrotron peak frequency ($\nu_{\rm s}$) and peak flux ($\nu_{\rm s} f_{\nu_{\rm s}}$) as well as other model parameters. We let $\nu_{\rm s}$ and $\nu_{\rm s} f_{\nu_{\rm s}}$ as free parameters in stead of $N_0$ and $\gamma_{\rm b}$ (see equations (2), (5) in Zhang et al. 2009).
We use the minimization $\chi^2$ technique to perform the SED fits. The free parameter set of our SED modeling is $\{B$, $\delta, \nu_{\rm s}, \nu_{\rm s} f_{\nu_{\rm s}}, \gamma_{\min}\}$, where $B$ is the magnetic field strength of the radiating region. We randomly generate a parameter set in broad spaces and measure the consistency between the model result and the observation data with a probability $p\propto e^{-\chi^2_{\rm r}/2}$, where $\chi^2_{\rm r}$ is the reduced $\chi^2$. Note that no errors are available for some data points. We estimate their errors with the average relative errors of the data points whose errors are available. Since the relative errors of the data for the synchrotron bump and SSC bump are usually much smaller than that for the EC bump, we derive the average relative errors of the data for the two bumps separately (with a separation at $\nu=10^{20}$ Hz; e.g., Zhang et al. 2012). For PKS 1510-089, no error of observation data below $10^{20}$ Hz is available, and we take $10\%$ of the observation flux as the errors (Zhang et al. 2012). A significant bump at the ultraviolet band is observed in the SEDs of some sources and it may be the thermal radiation of the accretion disk. We do not include these data in our SED modeling.

\begin{figure*}
\includegraphics[angle=0,scale=0.56]{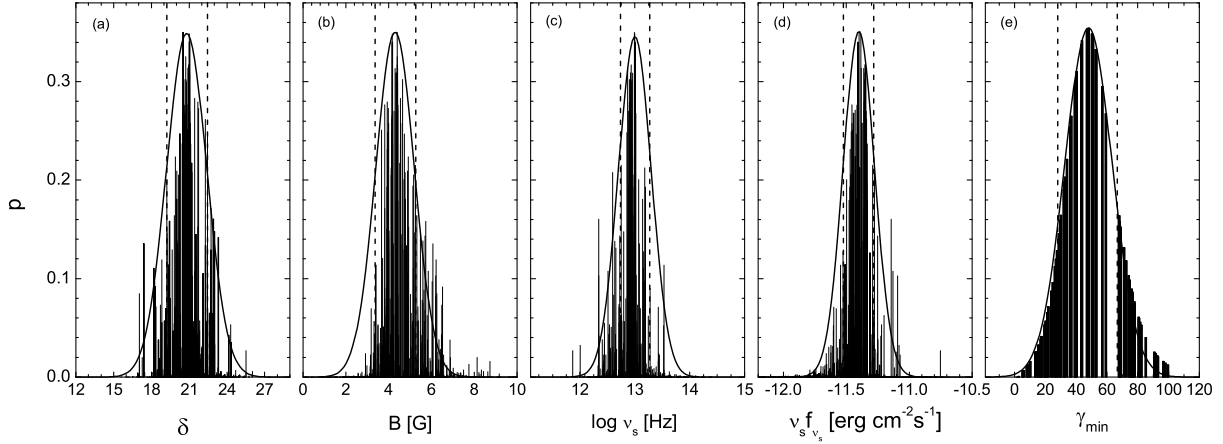}
\caption{Probability distributions of $\delta$, $B$, $\nu_{\rm s}$, $\nu_{\rm s}f_{\nu_{\rm s}}$, and $\gamma_{\rm min}$ for B2 1520+31. Gaussian function fits to the profiles of distributions are shown with {\em solid lines}. The vertical {\em dashed lines} mark the 1$\sigma$ ranges of the parameters.}\label{contour}
\end{figure*}

\begin{figure*}
\includegraphics[angle=0,scale=0.31]{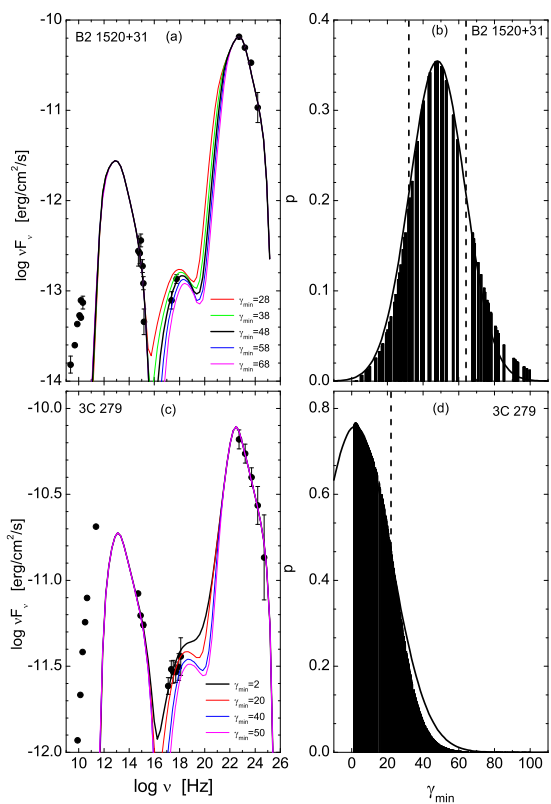}
\caption{Observed SEDs with model fitting lines at different $\gamma_{\rm min}$ values for B2 1520+31 ({\em panel a}) and 3C 279 ({\em panel c}), and the corresponding probability distributions of their $\gamma_{\rm min}$ with Gaussian function ({\em panel b}) and one-side Gaussian function ({\em panel d}) fits. The vertical {\em dashed lines} mark the 1$\sigma$ ranges of $\gamma_{\rm min}$.}\label{gam_min}
\end{figure*}

\begin{figure*}
\includegraphics[angle=0,scale=0.31]{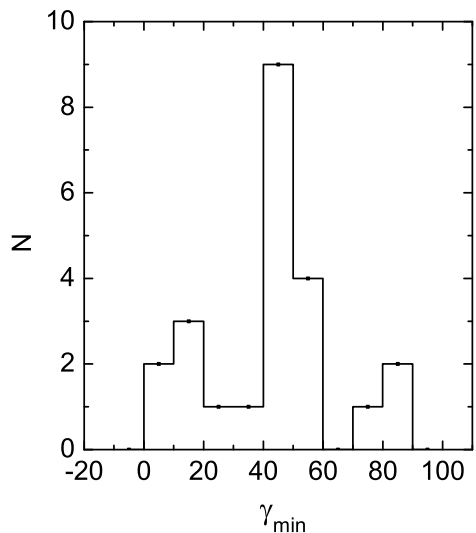}
\caption{Distribution of $\gamma_{\rm min}$ for the 23 FSRQs in our sample.}\label{gam_mindis}
\end{figure*}

Taking source B2 1520+31 as an example, Figure \ref{contour} shows the probability distributions of $B$, $\delta$, $\nu_s$, $\nu_s f_{\nu_s}$,
and $\gamma_{\min}$. The center values and $1\sigma$ confidence level of these parameters are derived from Gaussian fits to the profiles of the $p$
distributions. Note that the low energy data of the SSC bumps are sensitive to the $\gamma_{\min}$ values which have been reported by
Tavecchio et al. (2000) and Ghisellini et al. (2009b), as shown in Figures \ref{gam_min}(a) and (c). The probability distribution profiles of $\gamma_{\min}$ for the FSRQ SEDs in our sample are classified into two kinds as shown in Figure \ref{gam_min}(b) and (d). The $p$ profiles of five FSRQs (3C 279, 3C 273, PKS 0454-234, S4 0917+44, PKS 1502+106) are a one-side Gaussian function with a cutoff at $\gamma_{\min}=1$. This cutoff is due to the physical limit, i.e., $\gamma_{\min}$ should be $\geqslant 1$. Their $p$ profiles peak at $\gamma_{\min}=2$ or keep almost a constant in the range of $2\sim 20$, indicating that $\gamma_{\min}=2\sim 20$ can yield the best fit to the observation data. We thus fit them with a one-sided Gaussian function to derive the central values of $\gamma_{\min}$ and their upper limits in $1 \sigma$ confidence level, where we set their lower limits of $\gamma_{\min}$ at 2. We thus obtain $\gamma_{\min}=2^{+20}_{-0}$, $6^{+19}_{-4}$, $17^{+18}_{-15}$, $10^{+19}_{-8}$, $14^{+19}_{-12}$ for 3C 279, 3C 273, PKS 0454-234, S4 0917+44, PKS 1502+106, respectively. We show the distribution of $\gamma_{\min}$ for the given SEDs of FSRQs in our sample in Figure \ref{gam_mindis}. It ranges in 2$\sim$86 with a median of $\sim48$. Our SED fits are shown in Figure 1, and the model parameters are reported in Table 1 of Zhang et al. (2013). We also calculate the bolometric luminosities ($L_{\rm bol}$) based on our best SED fits. They are also reported in Table 2.

Note that blazars are violently variable. The model parameters derived from the fit to an SED are only for a given state of the sources (e.g., Zhang et al. 2012, 2013). The conventional method of the SED fits is through artificially adjusting the model parameters to make an acceptable fit. With the minimization $\chi^2$ technique, we not only can present a nice fit to the data, but also can yield the statistical confidence level of the parameters.  We should emphasize that the derived $1\sigma$ confidence level for the parameters with the minimization $\chi^2$ technique is not the observational errors of the parameters, but the confidence level in what extent the parameters can reproduce the observed SEDs.

\section{Jet Power and Cavity Kinetic Power}
The jet power ($P_{\rm jet}$) is essential to understand the production and composition of the jets. It is no way to directly measure the ratio of protons to electrons in the jets. The Proton-electron pair assumption\footnote{We further discuss this issue and examine another extreme case of the positron-electron pairs in Section 8.2.} is widely adopted in the calculations of jet power in blazars (e.g., Ghisellini et al. 2009b; 2010). We assume that the jet power is carried by relativistic electrons, cold protons, magnetic fields, and radiation, i.e., $P_{\rm jet}=\sum_i\pi R^2 \Gamma^2 c U^{'}_{\rm i}$, where $U^{'}_{i} (i={\rm e,\ p,}\ B, \rm r)$ are the energy densities associated with the emitting electrons
($U^{'}_{\rm e}$), cold protons ($U^{'}_{\rm p}$), magnetic field ($U^{'}_{B}$), and radiation\footnote{Note that the radiation power $P_{\rm r}$ should be a part of $P_{\rm jet}$ before radiation, since $P_{\rm e}$ is only the power carried by the electrons after radiation. We did not take the radiation power $P_{\rm r}$ into account when we calculated the total jet power $P_{\rm jet}$ in our previous work Zhang et al. (2012). Here we add the radiation power to the total jet powers of BL Lacs. } ($U^{'}_{\rm r}$) measured in the comoving frame (Ghisellini et al. 2010), which are given by
\begin{eqnarray}
U_{\rm e}^{'}=m_{\rm e}c^2\int N(\gamma)\gamma d\gamma,\\
U_{\rm p}^{'}=m_{\rm p}c^2\int N(\gamma)d\gamma,\\
U_B^{'}=B^2/8\pi,\\
U_{\rm r}^{'}=\frac{L_{\rm obs}}{4\pi R^2c\delta^4}\approx\frac{L_{\rm bol}}{4\pi R^2c\delta^4}.
\end{eqnarray}
We calculate the total jet powers and the powers carried by each components with our SED fitting parameters. The results are reported in Table 2.

The jet power also can be derived from the lobe low frequency radio emission under the assumption of minimum energy arguments
(e.g., Rawlings \& Saunders 1991; Willott et al. 1999). This approach now is widely used to estimate the jet kinetic energy in AGNs. It is also  believed that the observed X-ray cavities are evidence for AGN feedback and provide a direct measurement of the mechanical energy released by AGNs (B\^{\i}rzan et al. 2008; Cavagnolo et al. 2010). It is found that the cavity kinetic power is correlated with the radio extended power of galaxies (B\^{\i}rzan et al. 2004; 2008; Cavagnolo et al. 2010; O'Sullivan et al. 2011), and the scaling relationship is roughly consistent with the theoretical relation presented in Willott et al. (1991) (Cavagnolo et al. 2010; Meyer et al. 2011). We thus also estimate the cavity kinetic power $L_{\rm kin}$ using the relation between $L_{\rm kin}$ and radio luminosity (Meyer et al. 2011), i.e.,
\begin{equation}
\log L_{\rm kin}=0.64(\pm0.09)(\log L_{300}-40)+43.54(\pm0.12)   \quad(\rm erg/s),
\end{equation}
where $L_{300}$ is the extended luminosity at 300 MHz. Since the values of $L_{300}$ are not available for the sources in our sample. We estimated their values with the 5 GHz core luminosity. Meyer et al. (2011) defined a core dominance parameter with $R_{\rm CE}=\log (L_{\rm core}/L_{\rm ext})$ at 1.4 GHz. As shown in Figure 5 of Meyer et al. (2011), the typical $R_{\rm CE}$ for blazars is 0.5, which is adopted in our calculations, i.e., $L_{\rm core}/L_{\rm ext}=3$. We first use the spectral index of $\alpha=0.5$ (radio spectral index for the core region, Urry \& Padovani 1995) to derive the core luminosity at 1.4 GHz, and then employ the core dominance parameter $R_{\rm CE}$ to obtain the extended luminosity at 1.4 GHz for the FSRQs and BL Lacs in our samples. We finally obtain $L_{300}$ using a radio spectral index of $\alpha=1.2$ (radio spectral index for extended region, Meyer et al. 2011). We also derive the errors of $L_{\rm kin}$ by considering the uncertainty of the correlation in equation (5).

\begin{figure*}
\includegraphics[angle=0,scale=0.56]{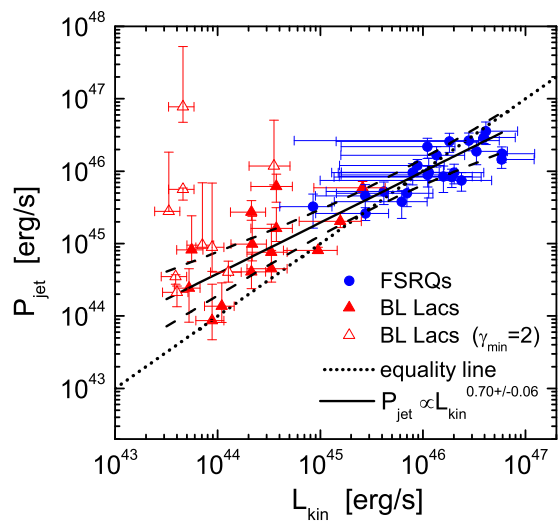}
\caption{Comparison between $P_{\rm jet}$ with $L_{\rm kin}$ for FSRQs and BL Lacs. The {\em solid line} is the best fits for FSRQs
and BL Lacs (without considering the opened triangles data points). The {\em dashed lines} indicate the 3$\sigma$ confidence bands for the best fits. The {\em dotted line} is the equality line. }\label{Lkin}
\end{figure*}

Comparison between $P_{\rm jet}$ and $L_{\rm kin}$ for the FSRQs and BL Lacs in our sample is shown in Figure \ref{Lkin}. Note that the values of $\gamma_{\min}$ for some BL Lacs are poorly constrained by the observed SEDs (Zhang et al. 2012). We took $\gamma_{\min}=2$ in the calculations of $P_{\rm e}$ and $P_{\rm p}$ for those BL Lacs whose $\gamma_{\min}$ values cannot be constrained with the data (Zhang et al. 2012). This may lead to significant over-estimate of their $P_{\rm jet}$, mainly $P_{\rm p}$, if their true $\gamma_{\min}$ values are much larger than 2. We mark these sources as different symbols in the figures, but do not take these sources (10 data points) into account for statistical analysis hereafter. One can observe that $P_{\rm jet}$ and $L_{\rm kin}$ are correlated. The best linear fit yields $\log P_{\rm jet}=(13.7\pm2.9)+(0.70\pm0.06)\log L_{\rm kin}$, with a Pearson correlation coefficient $r=0.90$ and a chance probability $p=3.9\times10^{-14}$.

\section{Jet Composition and Radiation Efficiency}
\begin{figure*}
\includegraphics[angle=0,scale=0.6]{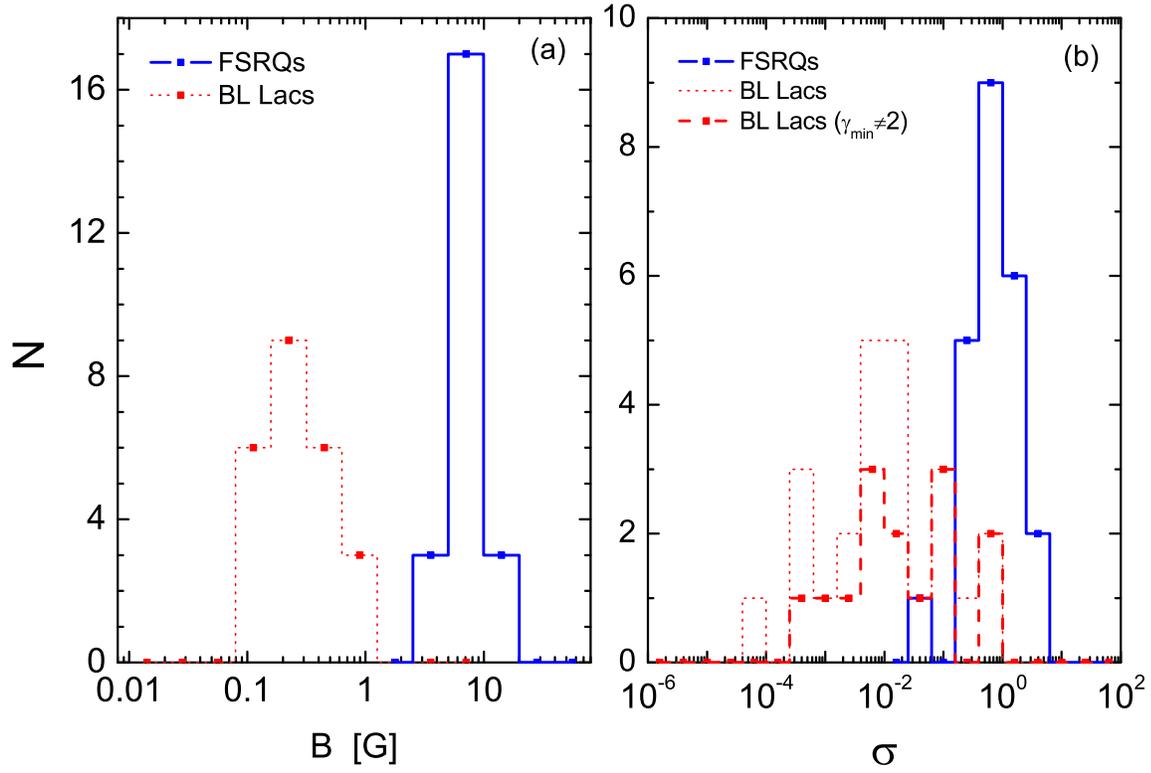}
\caption{Comparisons of the magnetic field strength ({\em panel a}) and the magnetization parameter ({\em panel b}) between the FSRQs ({\em blue solid lines}) and BL Lacs ({\em red thin dotted lines} for all the sources and {\em red thick dashed line} for sources with $\gamma_{\min}\neq2$) in our sample.}\label{B_Dis}
\end{figure*}

\begin{figure*}
\includegraphics[width=2.in,height=7.5in]{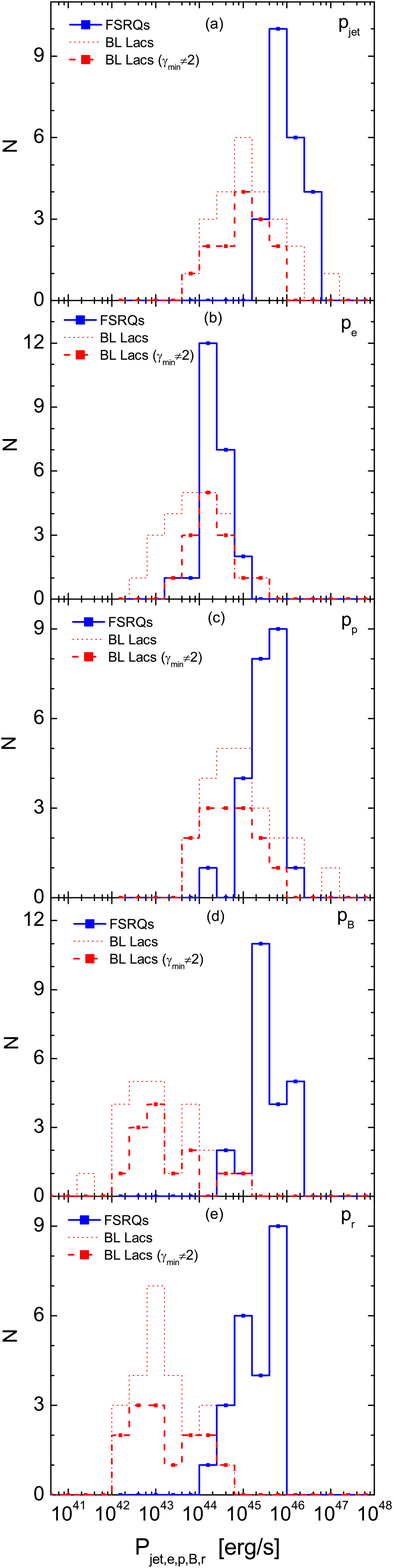}
\includegraphics[width=2.in,height=6.1in]{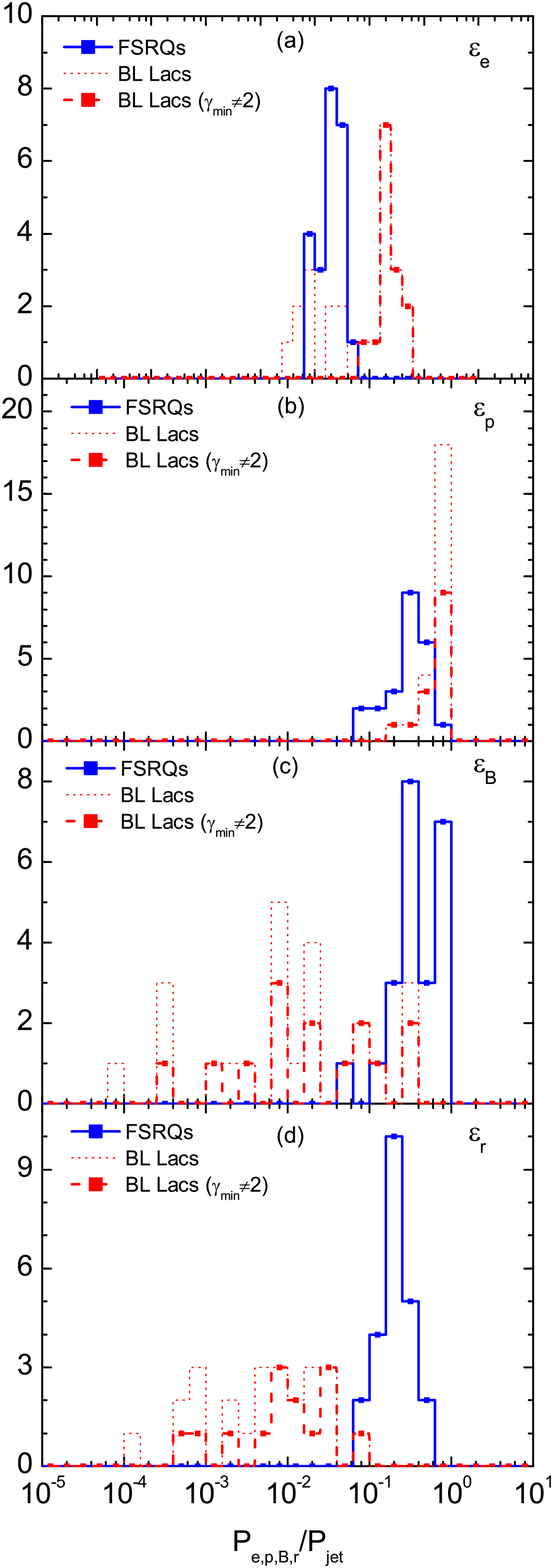}
\caption{Distributions of the powers associated with relativistic electrons $P_{\rm e}$, cold protons $P_{\rm p}$,
Poynting flux $P_B$, radiation component $P_{\rm r}$, and the total power $P_{\rm jet}$ of the jets ({\em left panels}),
together with the distributions of the ratios of these powers to $P_{\rm jet}$ ({\em right panels}). The line styles are the same as in Fig. \ref{B_Dis}.}\label{Pjet}
\end{figure*}

Whether the jets are matter or Poynting flux dominating is essential to understand the radiation physics and jet formation of blazars. We first compare the distributions of $B$ between the two kinds of sources in Figure \ref{B_Dis}(a). The magnetic field strengths of the FSRQs are larger than that of the BL Lacs by 1-2 orders of magnitude, i.e., from 3.1 G to 15.1 G for FSRQs and from 0.09 G to 1.1 G for BL Lacs. The medians of the distributions are 6.9 G for FSRQs and 0.2 G for BL Lacs. We then compare the powers carried by the jet ingredients for the FSRQs and BL Lacs in Figure \ref{Pjet}. One can find that the jet powers of FSRQs are systematically larger than that of BL Lacs. The values of $P_{\rm jet}$ for the FSRQs are from $2.6\times10^{45}$ erg $\rm s^{-1}$ to $3.6\times10^{46}$ erg $\rm s^{-1}$ with a median of $9.6\times10^{45}$ erg $\rm s^{-1}$, and the median is $8.1\times10^{44}$ erg $\rm s^{-1}$ for the BL Lacs. Similarly, the medians of the $P_{\rm e}$ and $P_{\rm p}$ distributions are $2.0\times10^{44}$ erg $\rm s^{-1}$ and $3.0\times10^{45}$ erg $\rm s^{-1}$ for FSRQs, whereas they are $1.9\times10^{44}$ erg $\rm s^{-1}$ and $5.1\times10^{44}$ erg $\rm s^{-1}$ for BL Lacs, respectively. However, the medians of the $P_{B}$ and $P_{\rm r}$ distributions for FSRQs are much larger than that for BL Lacs with almost three orders of magnitude, i.e., $3.0\times10^{45}$ erg $\rm s^{-1}$ and $2.4\times10^{45}$ erg $\rm s^{-1}$ for FSRQs and $1.1\times10^{43}$ erg $\rm s^{-1}$ and $1.3\times10^{43}$ erg $\rm s^{-1}$ for BL Lacs, respectively.

\begin{figure*}
\includegraphics[angle=0,scale=0.23]{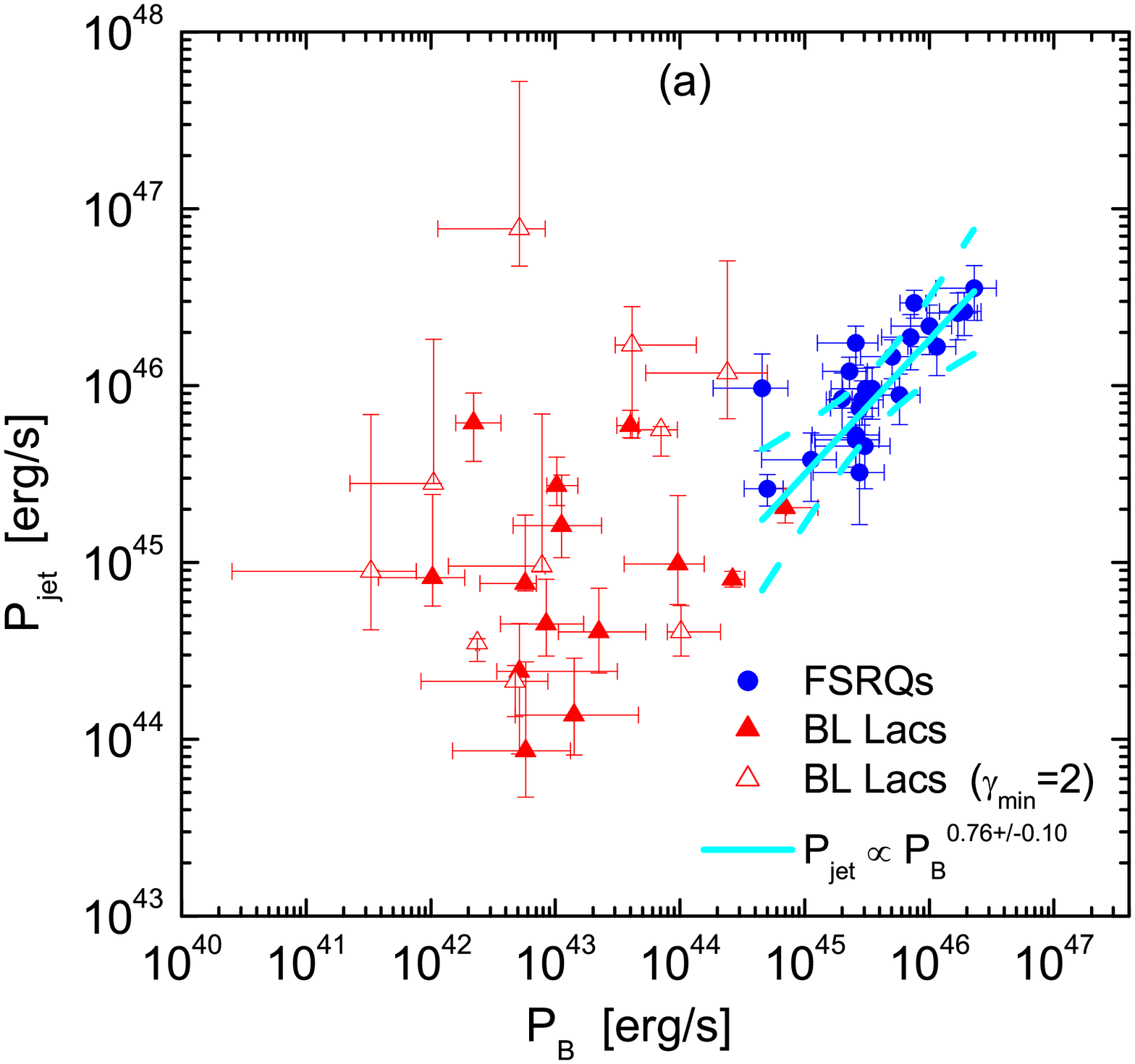}
\includegraphics[angle=0,scale=0.23]{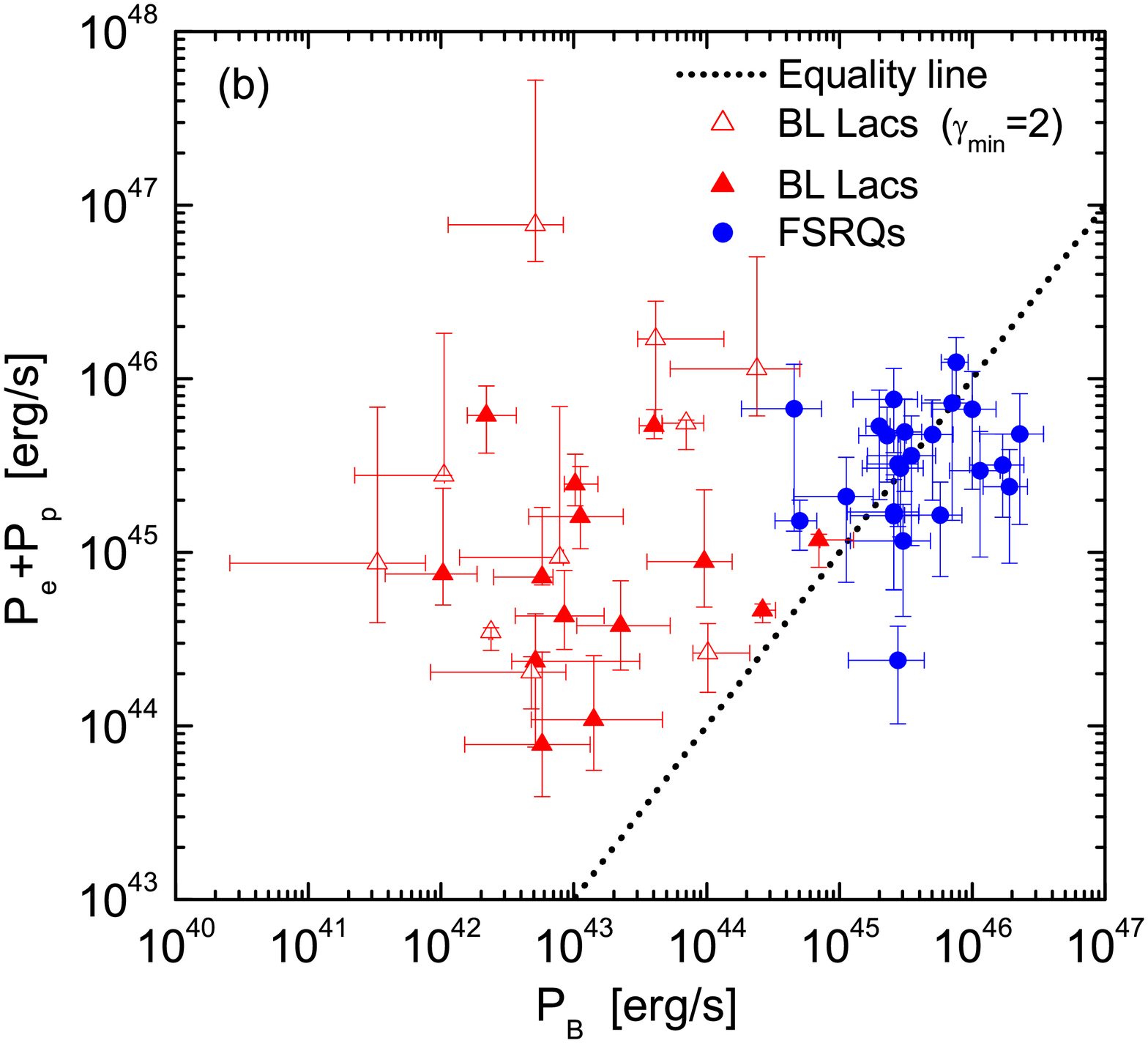}
\includegraphics[angle=0,scale=0.23]{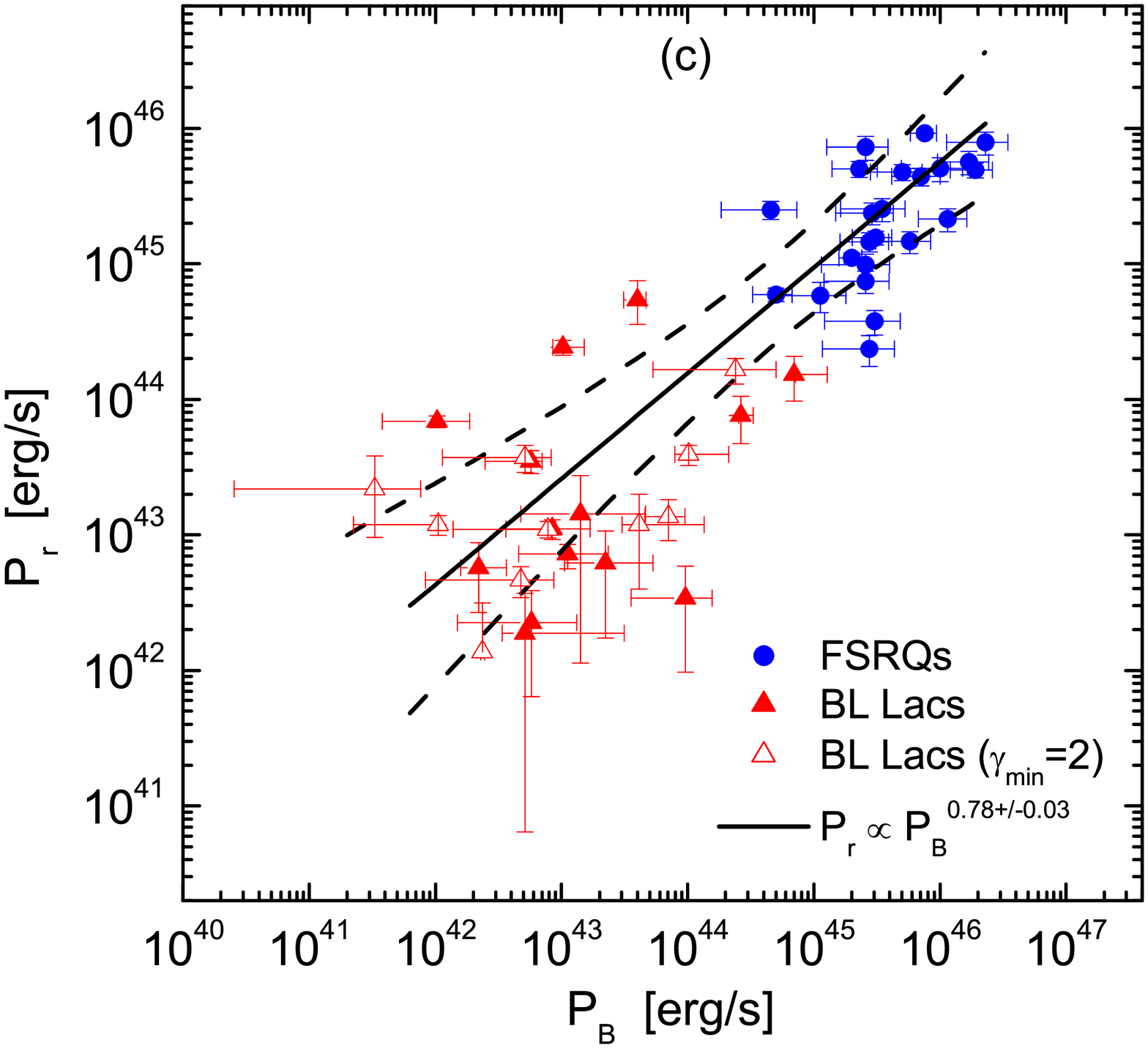}
\caption{$P_{\rm jet}$, $P_{\rm e}+P_{\rm p}$, and $P_{\rm r}$ as a function of $P_{\rm B}$ for FSRQs and BL Lacs. The {\em solid lines} are the best fits and the {\em dashed lines} indicate the 3$\sigma$ confidence bands for the best fits. The {\em dotted line} in {\em panel b} is the equality line.} \label{Pjet_Pb}
\end{figure*}

We calculate the ratios of the power carried by each ingredient to $P_{\rm jet}$, i.e., $\epsilon_{\rm i}=P_{\rm i}/P_{\rm jet}$, and show their distributions in Figure \ref{Pjet}. It is found that the medians of $\epsilon_{\rm e}$ and $\epsilon_{\rm p}$ distributions for FSRQs are 0.02 and 0.34, respectively, being smaller than that of the BL Lacs, which are 0.24 and 0.67, respectively. However, the medians of the $\epsilon_{B}$ and $\epsilon_{\rm r}$ distributions for FSRQs are 0.37 and 0.22, being much larger than that of BL Lacs, which are 0.02 and 0.04, respectively. In addition, as shown in Figure \ref{Pjet_Pb}(a) $P_{\rm jet}$ of FSRQs is tightly correlated with $P_{B}$, i.e., $P_{\rm jet}\propto P_{B}^{0.76\pm0.10}$ with a Pearson correlation coefficient of $r=0.77$ and chance probability of $p=1.8\times10^{-5}$; however it is not so for the BL Lacs. We also show $P_{\rm e}+P_{\rm p}$ as a function of $P_{B}$ in Figure \ref{Pjet_Pb}(b). One can observe that $P_{\rm B}$ is larger than $P_{\rm e}+P_{\rm p}$ for half of FSRQs, implying that the jets of FSRQs may be dominated by Poynting flux, but is not for the case of BL Lacs, which may be dominated by particles.

We measure the magnetization of a jet with a parameter $\sigma=P_{B}/(P_{\rm p}+P_{\rm e}+P_{\rm r}$) (Zhang et al. 2013). We show the distributions of $\sigma$ for FSRQs and BL Lacs in Figure \ref{B_Dis}(b). One can observe that the FSRQs tend to have a higher $\sigma$ than the BL Lacs, sometimes even close to or exceeding unity. The radiation efficiencies $\epsilon_{\rm r}$ for most of the FSRQs are greater than 0.1, and even close to 0.5 as shown in Figure \ref{Pjet}. For BL Lacs, $\epsilon_{\rm r}$ distributes in a very broad range, but always smaller than 0.1 and sometimes even lower than $10^{-3}$. As reported in Zhang et al. (2013), the FSRQ jets are likely highly magnetized and the BL Lac jets have lower radiation efficiency and are matter dominating. The radiation power of jet $P_{\rm r}$ as a function of $P_{B}$ is shown in Figure \ref{Pjet_Pb}(c). FSRQs and BL Lacs form a clear sequence that a jet with higher $P_{B}$ tends to have higher $P_{\rm r}$. The best linear fit in log scale gives $P_{\rm r} \propto P_{B}^{0.78\pm 0.03}$ with a Pearson correlation coefficient of $r=0.86$ and chance probability of $p=7.3\times 10^{-12}$. It seems that a jet with higher $P_{B}$ tends to be much tightly correlated with its $P_{\rm r}$. These results indicate that the high radiation efficiency of FSRQ jets could be due to their high magnetization of jets.

\section{Jet Bulk Motion and Electron Spectrum}
\begin{figure*}
\includegraphics[angle=0,scale=0.25]{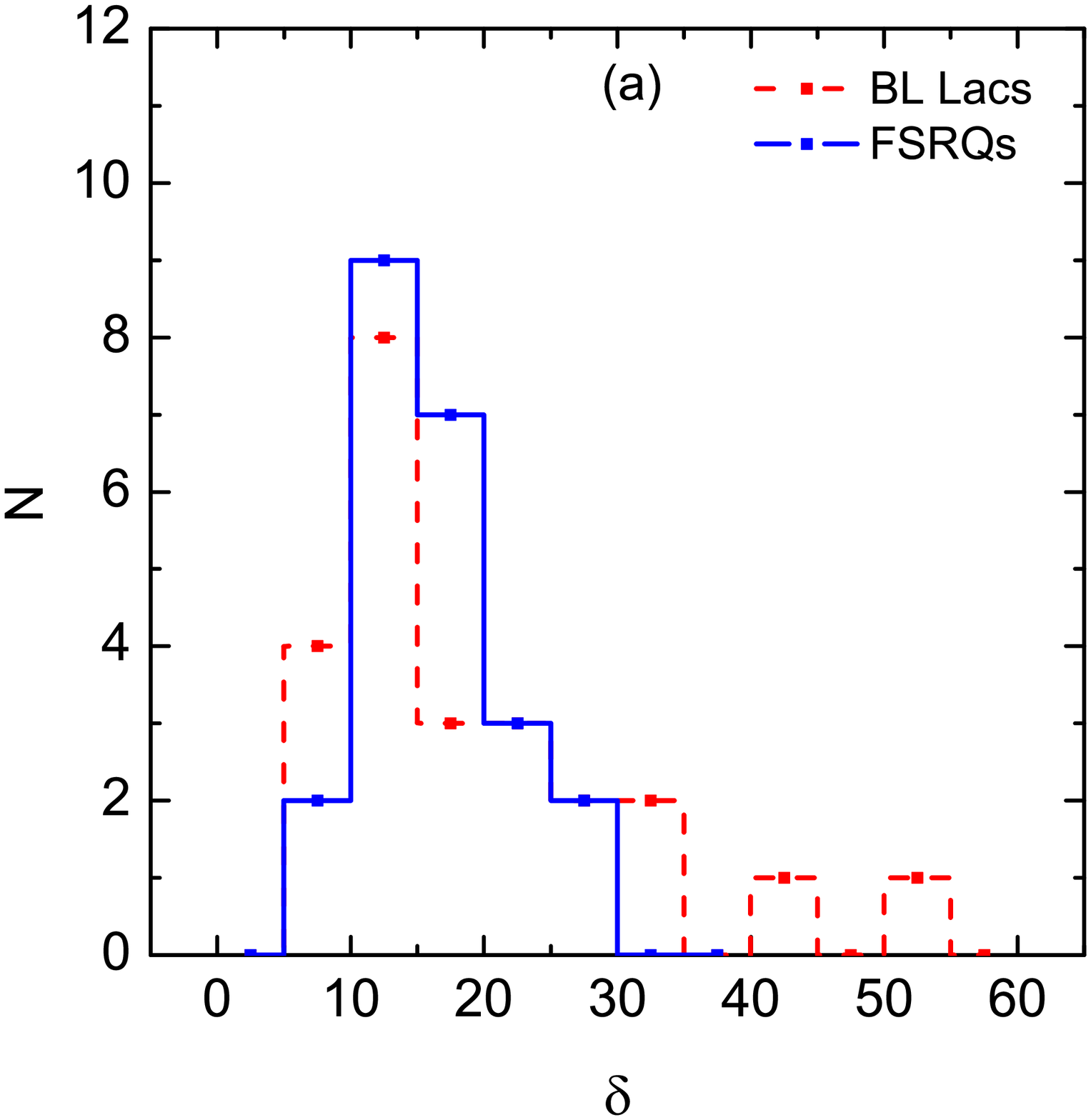}
\includegraphics[angle=0,scale=0.25]{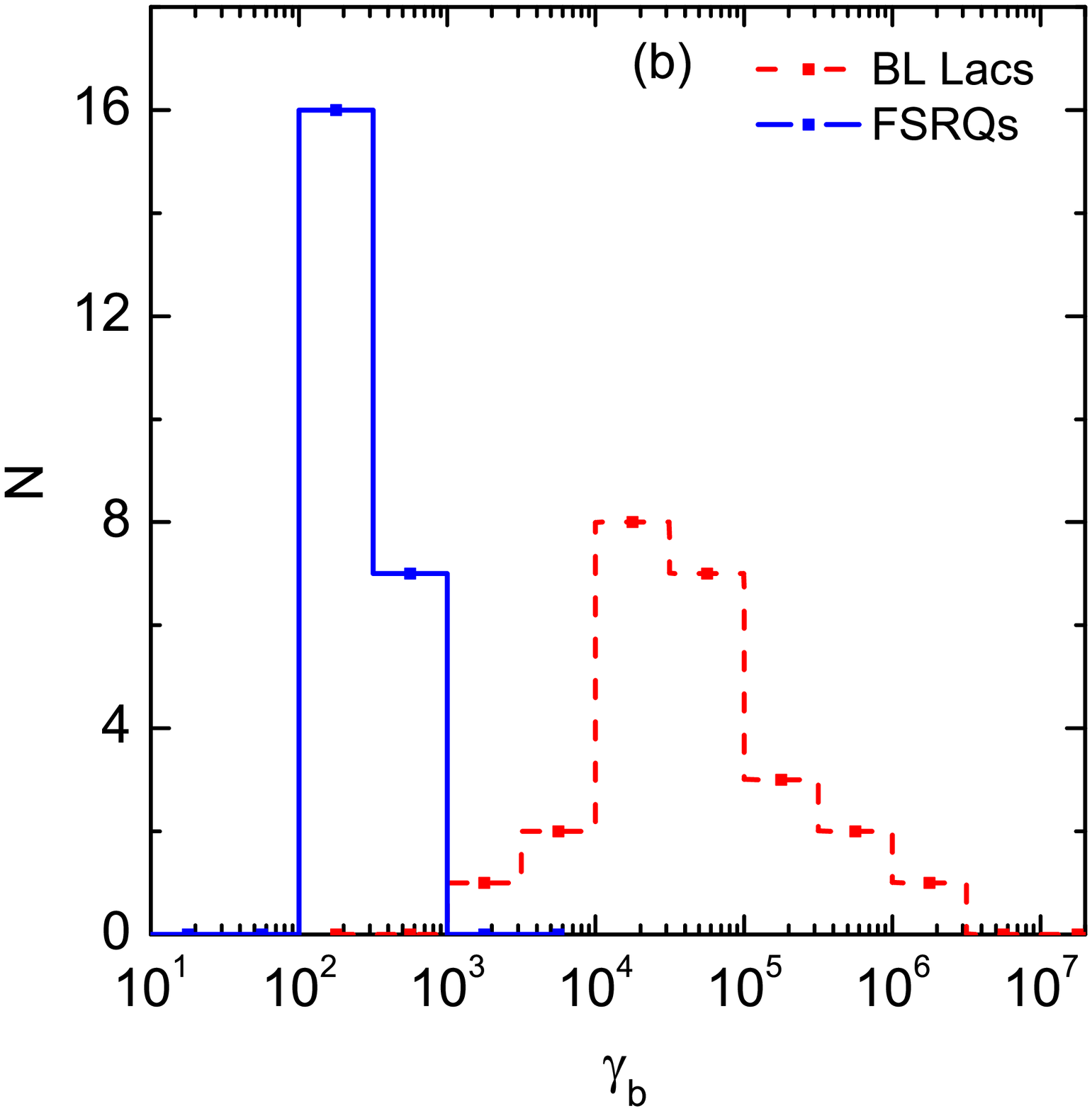}
\includegraphics[angle=0,scale=0.25]{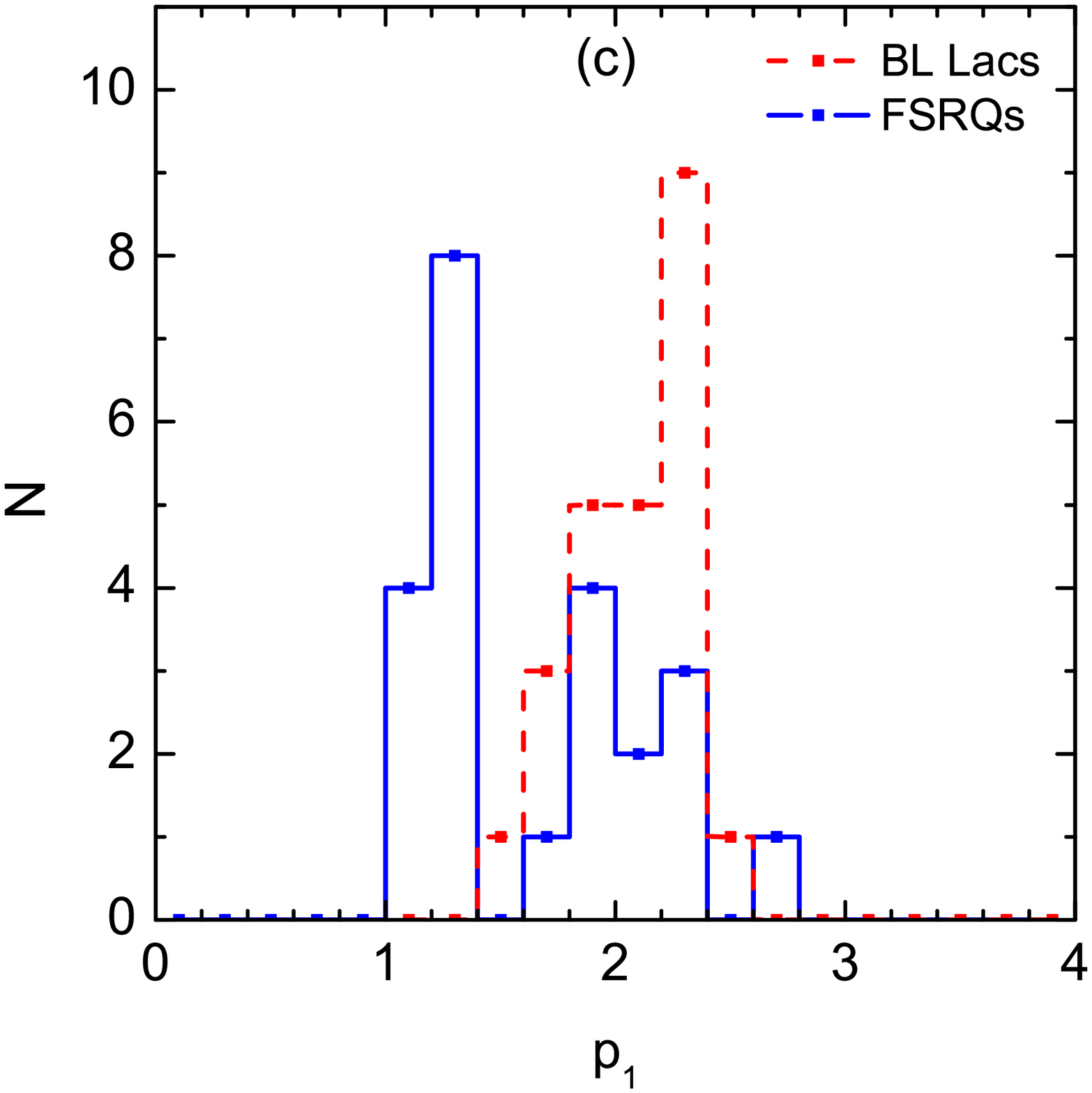}
\includegraphics[angle=0,scale=0.25]{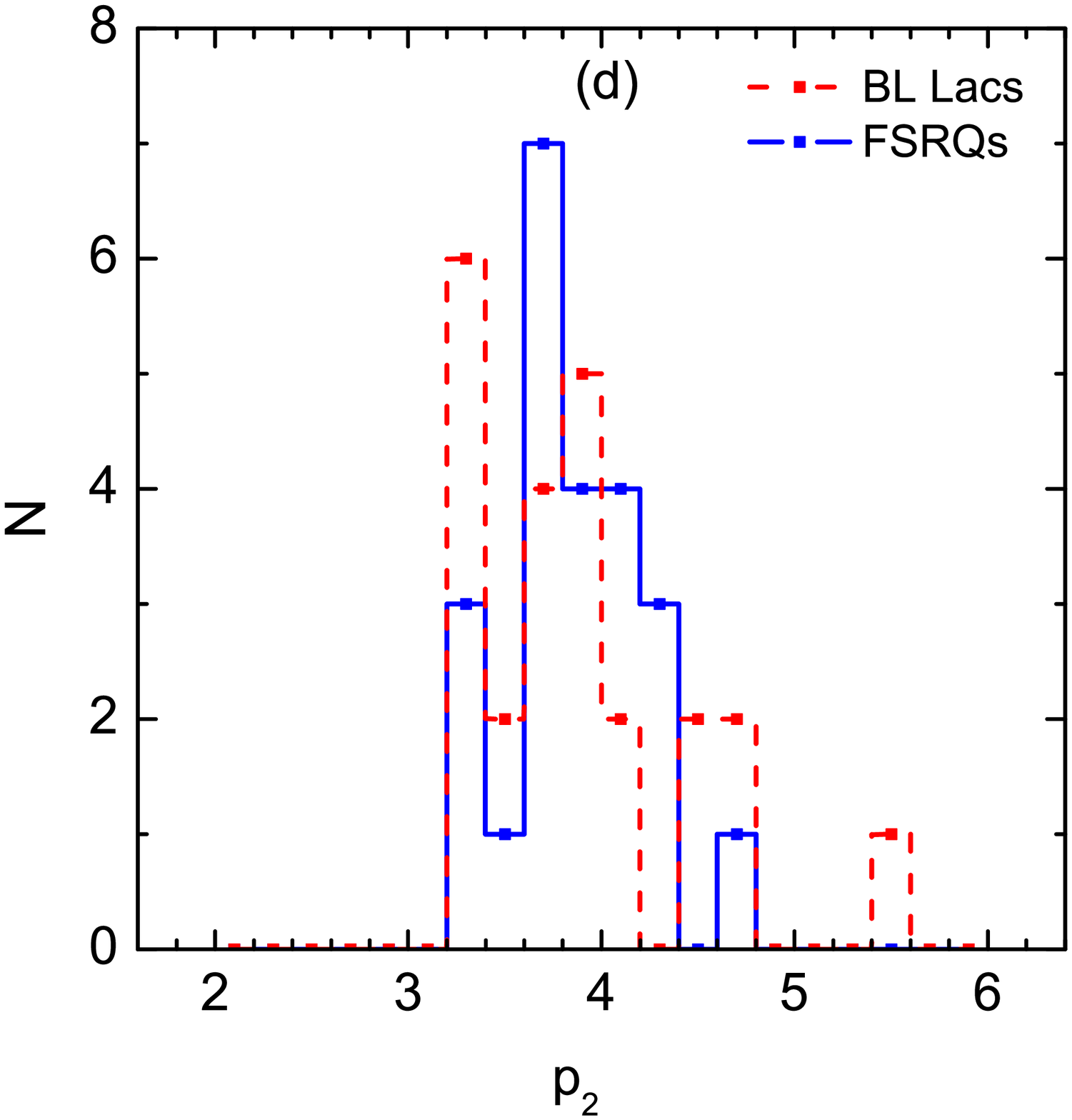}
\hfill
\includegraphics[angle=0,scale=0.25]{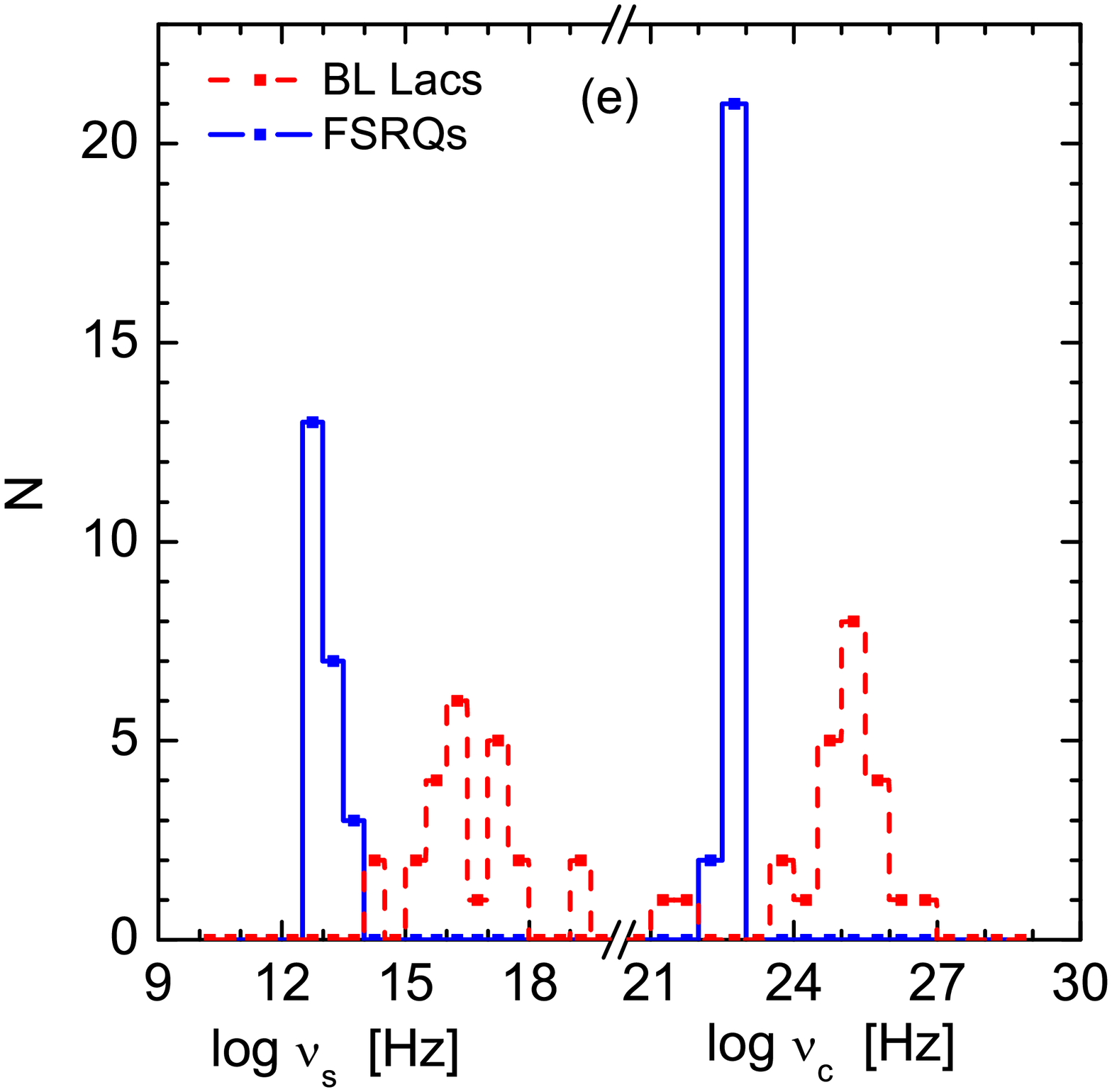}
\hfill
\includegraphics[angle=0,scale=0.25]{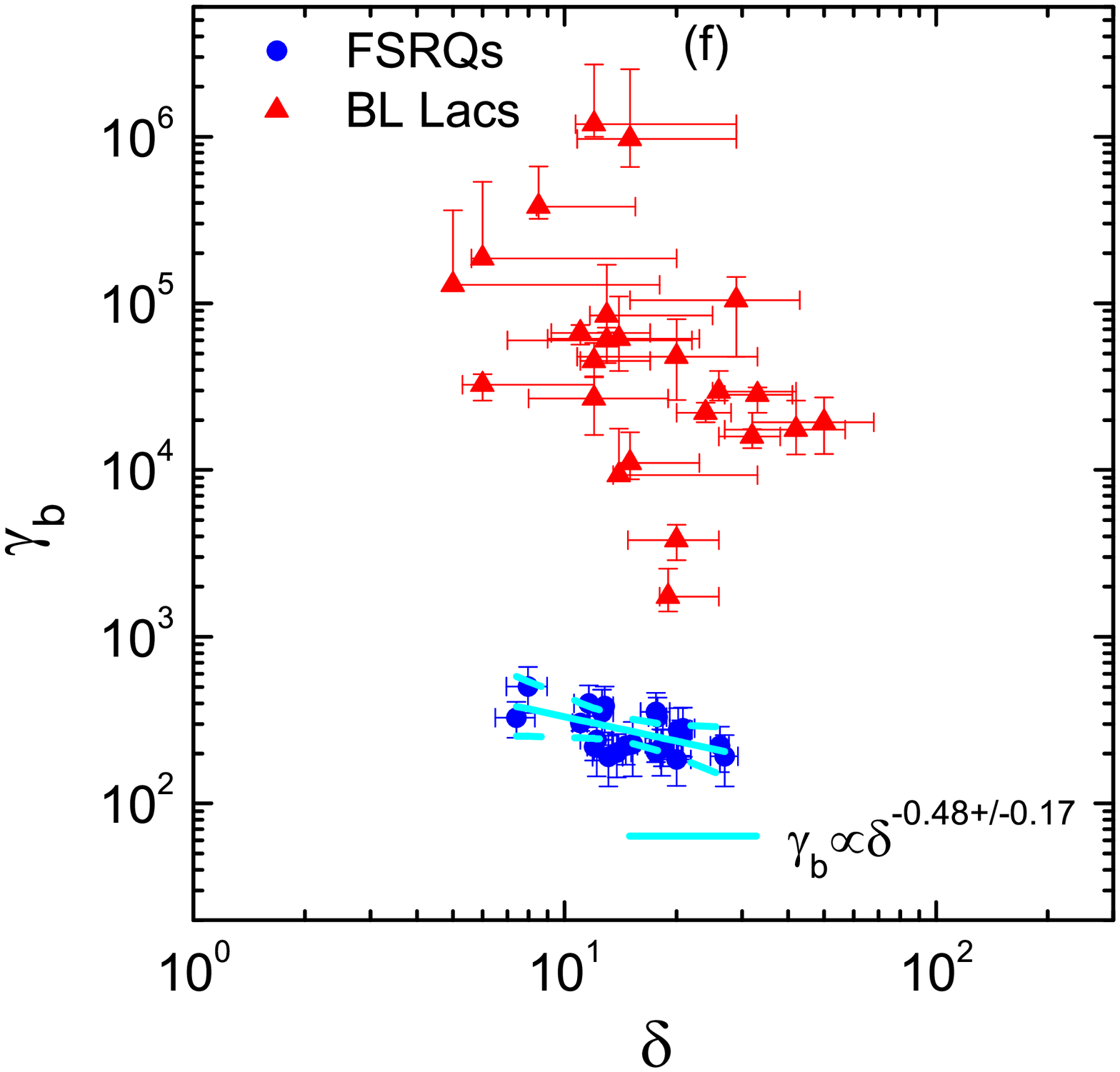}
\caption{ Distributions of the beaming factor $\delta$, the break Lorentz factor of electrons $\gamma_{\rm b}$, the indices $p_1$ and $p_2$ of the electron distribution, and the two peak frequencies $\nu_{\rm s}$ and $\nu_{\rm c}$ for the 23 FSRQs with the BL Lac data from Zhang et al. (2012). {\em Panel f} --- $\gamma_{\rm b}$ as a function of $\delta$. The {\em solid line} is the best fit for FSRQ data and the {\em dashed lines} indicate the 3$\sigma$ confidence bands of the best fit.}.\label{Electron_Dis}
\end{figure*}

We compare the distributions of $\delta$ and the parameters of the electron spectrum between FSRQs and BL Lacs in Figure \ref{Electron_Dis}. It is found that $\delta$ for FSRQs ranges from 10 to 27 with a same median of $\sim15$ as BL Lacs. We test whether the two distributions show any statistical difference with the Kolmogorov-Smirnov test (K-S test), which yields a chance probability $p_{\rm KS}$. A K-S test probability larger than 0.1 would strongly suggest no statistical difference between two distributions. We get $p_{\rm KS}=0.63$, indicating that the distributions of $\delta$ for the two kinds of blazars have no statistical difference. We do not find any relation between $\delta$ and $B$, hence the magnetic field is not related to the bulk motion of the emitting regions for both FSRQs and BL Lacs.

The electron spectrum may signal the acceleration of the electrons in the radiation region. We also compare the parameter distributions of the electron spectra and the two peak frequencies of SEDs between the FSRQs and BL Lacs in Figures \ref{Electron_Dis}(b), (c), (d), (e). The distributions of $p_2$ for the two kinds of blazars\footnote{The K-S tests yield $p_{\rm KS}=2\times10^{-3}$ and $p_{\rm KS}=0.51$ for the distributions of $p_{1}$ and $p_{2}$ between FSRQs and BL Lacs.} are consistent with the same mean of $p_2=3.8$. $p_1$ for BL Lacs is around 2, roughly consistent with the prediction of Fermi acceleration mechanism (Gallant 2002; Wang 2002; Cao \& Wang 2013b; Zhou et al. 2013). The distribution of $p_1$ for FSRQs is bimodal; half are consistent with that for BL Lacs and half have $p_1\sim 1.2$, implying the different acceleration process (Yan et al. 2013). The electron acceleration is likely not related to the bulk motion of the jets since no statistical difference is observed in the $\delta$ distributions between FSRQs and BL Lacs, but the distributions of $\gamma_{\rm b}$ are dramatically different for FSRQs and BL Lacs. $\gamma_{\rm b}$ ranges from $10^{3}$ to $10^{6}$ for BL Lacs. However $\gamma_{\rm b}$ narrowly clusters around several hundreds for FSRQs. This results in very narrow distributions of both $\nu_{\rm s}$ and $\nu_{\rm c}$, i.e., $\nu_s=10^{12}\sim10^{13}$ Hz and $\nu_c=10^{22}\sim10^{23}$ Hz, as shown in Figure \ref{Electron_Dis}(e), which are also consistent with the results of Senturk et al. (2013). Note that the IC bumps of the SEDs for FSRQs are dominated by the EC process and the energy of seed photons is $2\times10^{15}\Gamma$ Hz in the comoving frame. The peak frequencies of EC bumps are $\nu_{\rm c}\sim 2\times10^{15}\Gamma\gamma_{\rm b}^2\delta$. We find a tentative anti-correlation between $\gamma_{\rm b}$ and $\delta$ with $r=-0.54$ (Pearson correlation coefficient) and $p=0.008$ (chance probability) for FSRQs as shown in Figure \ref{Electron_Dis}(f). The narrow distributions of $\gamma_{\rm b}$ and $\delta$ result in the narrow distribution of $\nu_{\rm c}$ for the FSRQs. This anti-correlation may also imply that the relativistic electrons in the emission region with a larger $\delta$ suffer more cooling by the EC process. Therefore, the lower $\gamma_{\rm b}$ in FSRQs may be due to the strong cooling of relativistic electrons by both the SSC and EC processes. Ghisellini et al. (2010) also found similar feature for bright {\em Fermi} blazars. The observed blazar spectral sequence (e.g., Fossati et al. 1998; Donato et al. 2001) should be due to the different electron cooling efficiency (Ghisellini et al. 1998; 2009b; 2010; Sbarrato et al. 2012) or the increasing importance of the external radiation field along with the sequence of HBLs-LBLs-FSRQs.

\section{Connection between the Jet Properties and the Central Black Holes}
\begin{figure*}
\includegraphics[angle=0,scale=0.3]{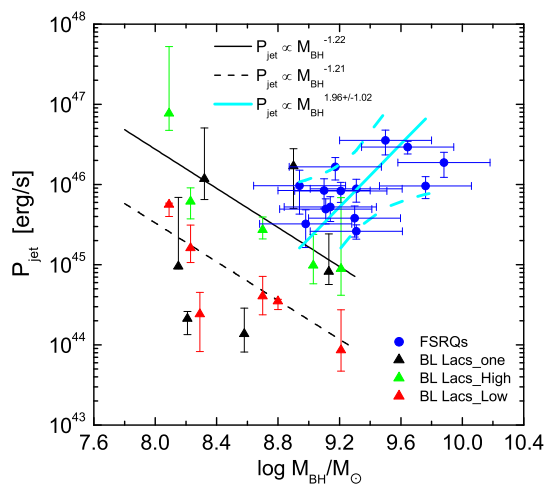}
\caption{Jet power as a function of the BH mass. The {\em triangles} indicate the BL Lacs. The {\em green and red triangles} are for the sources in the high and low states, respectively, and {\em black triangles} are for the sources with only one SED available as reported in Zhang et al. (2012). The {\em black solid and dashed lines} are fit lines for high and low states data and are also taken from Zhang et al. (2012). The {\em blue circles} indicate FSRQs.  The {\em cyan solid line} is the best fit for FSRQ data and the {\em cyan dashed lines} indicate the 3$\sigma$ confidence bands of the best fit.}\label{MbhPjet}
\end{figure*}

AGN jets may be driven by both the accretion process and the spin of the central BH (Fanidakis et al. 2011; Zhang et al. 2012). We further investigate the connection between the jet properties and the central BHs. We search the BH masses from literature and obtain a sub-sample of 14 FSRQs with BH masses available, as reported in Table 2. Although the BH masses for FSRQs are collected from different papers, thirteen FSRQs out of them were obtained using the same assumption that broad-line clouds are virialized. A sub-sample of BL Lacs with BH mass available is from Zhang et al. (2012). As reported in Zhang et al. (2012), a weak relation of $P_{\rm jet}\propto M_{\rm BH}^{-1.2}$ was found in both the high and low states of BL Lacs. We also examine the $P_{\rm jet}$-$M_{\rm BH}$ relation for FSRQs. Dramatically different from BL Lacs, a tentative positive correlation between $P_{\rm jet}$ and $M_{\rm BH}$ is found for FSRQs as shown in Figure \ref{MbhPjet}. The best linear fit in log scale gives $P_{\rm jet}\propto M_{\rm BH}^{1.96\pm 1.02}$, with $r=0.53$ and $p=0.05$, where the errors of BH masses are assumed to be $\triangle(\log M_{\bigodot})=0.3$ for all the FSRQs in our sample.

\begin{figure*}
\includegraphics[angle=0,scale=0.3]{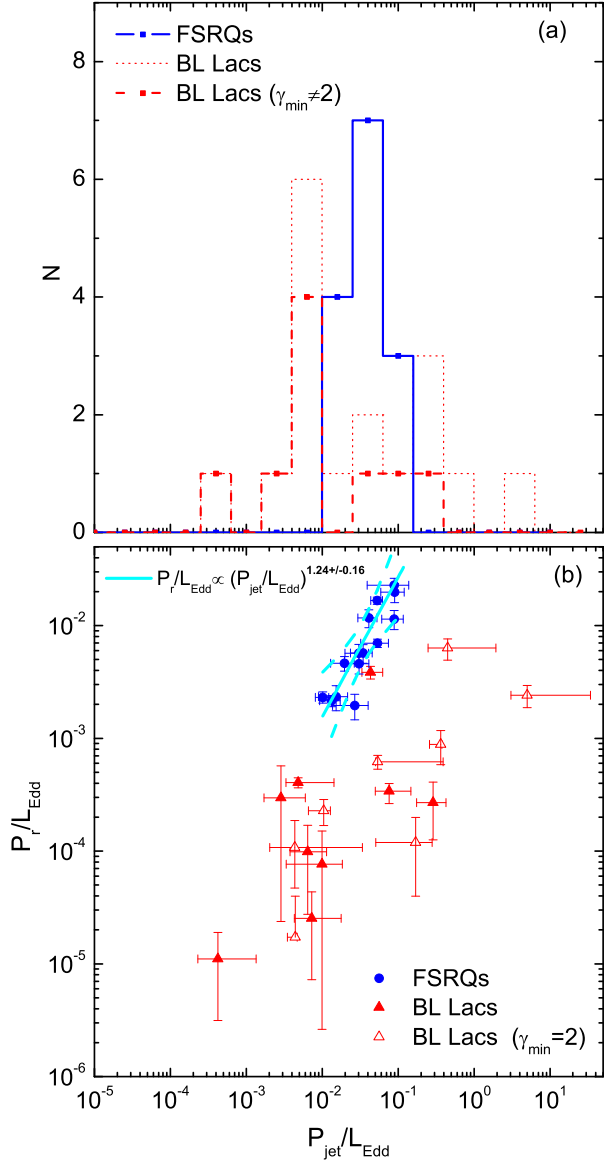}
\caption{{\em Panel a} --- Distribution of $P_{\rm jet}/L_{\rm Edd}$, the line styles are the same as in Fig. \ref{B_Dis}. {\em Panel b} --- $P_{\rm r}/L_{\rm Edd}$ as a function of $P_{\rm jet}/L_{\rm Edd}$ for FSRQs and BL Lacs in our sample. The {\em solid line} is the best fit
to the data of FSRQs and the {\em dashed lines} mark the corresponding 3$\sigma$ confidence bands of the best fit.}\label{MbhPjetLedd}
\end{figure*}

We measure the jet production and radiation rates per central BH mass with $P_{\rm jet}/L_{\rm Edd}$ and $P_{\rm r}/L_{\rm Edd}$, where $L_{\rm Edd}$ is the Eddington luminosity. Interestingly, as shown in Figure \ref{MbhPjetLedd}, $P_{\rm jet}/L_{\rm Edd}$ and $P_{\rm r}/L_{\rm Edd}$ of FSRQs narrowly range in $0.01\sim0.09$ and $0.002\sim0.023$, with averages of $<P_{\rm jet}/L_{\rm Edd}>=0.04$ and $<P_{\rm r}/L_{\rm Edd}>=0.008$, respectively. This likely indicates the universal jet production and radiation rates per central BH mass for FSRQs, hence the same dominating jet formation mechanism may work at these FSRQs. As shown in Figure \ref{MbhPjetLedd}(b), $P_{\rm r}/L_{\rm Edd}$ is tightly correlated with $P_{\rm jet}/L_{\rm Edd}$ for FSRQs, i.e., $P_{\rm r}/L_{\rm Edd}\propto (P_{\rm jet}/L_{\rm Edd})^{1.24\pm 0.16}$ derived from the best linear fit in log scale with a Pearson correlation coefficient of $r=0.89$ and a chance probability of $p=1.8\times10^{-5}$. For the BL Lacs, both $P_{\rm jet}/L_{\rm Edd}$ and $P_{\rm r}/L_{\rm Edd}$ span for several orders of magnitude (without considering data points of $\gamma_{\min}=2$), i.e., $P_{\rm jet}/L_{\rm Edd}=4\times10^{-4}\sim0.3$ and $P_{\rm r}/L_{\rm Edd}=10^{-5}\sim0.004$. As pointed out in Section 4, since the observed SEDs of some BL Lacs cannot present constraint on the minimum Lorentz factor of electrons, we take $\gamma_{\min}=2$ for those sources, which may lead to significant over-estimate of their $P_{\rm jet}$ and the values of $P_{\rm jet}/L_{\rm Edd}$ larger than unity. These results may imply that the dominating mechanisms of jet production in FSRQs and BL Lacs are different.

\section{Discussion}
\subsection{Jet Formation Mechanisms}
It is generally believed that the jet formation is via either the BP (Blandford \& Payne 1982) and/or BZ (Blandford \& Znajek 1977) mechanisms.   Besides the BH spin and the accretion rate, the dramatically different $P_{\rm jet}-M_{\rm BH}$ relations shown in Figure \ref{MbhPjet} between the FSRQs and BL Lacs may also signal that the mass of BH would be also an essential factor for the jet radiation efficiency and jet power (see also Davis \& Laor, 2011).

The different $P_{\rm jet}-M_{\rm BH}$ relations may indicate the different dominating jet formation mechanisms in the two kinds of blazars. The anti-correlation between $P_{\rm jet}$ and $M_{\rm BH}$ observed in the BL Lacs may disfavor the scenario that the jets are purely powered by accretion process and imply that the spin energy release of the central BH may also play a significant role in jet formation for these objects (Zhang et al. 2012). The mix of the two contributions may lead to a broad jet production rate among sources, hence there is no universal jet production efficiency among BL Lacs. This is supported by the broad $P_{\rm jet}/L_{\rm Edd}$ distribution of BL Lacs, which spans from $10^{-4}$ to 0.3. It was proposed that the different structures and accretion rates of accretion disks may result in the different dominating mechanisms of jet launching (Ghisellini \& Celotti 2001; Pu et al. 2012; Zhang 2013).  The accretion disk of a BL Lac with low accretion rate may be consisted of an outer thin disk and an inner advection-dominated accretion flow (e.g., Zhang 2013). In such a combined disk, which has a large inner radius, the accreted materials in the accretion disk would be ejected along with the line of magnetic force, and a relativistic jet with moderate luminosity and speed would be launched by the BP process. When the accretion rate and the BH spin increase, the inner radius of the disk becomes smaller and is closer to the BH. The dominating jet formation mechanism then may transfer to the BZ mechanism. Therefore, it is possible that there is no universal jet production mechanism at work among these BL Lacs.

The highly magnetized jets in FSRQs may be powered by the BZ mechanism. It was suggested that the magnetic flux threading the BH is important to launch a powerful jet (Sikora \& Begelman et al. 2013).
The efficiency of extracting the BH rotational energy via the BZ mechanism depends on the magnetic flux dragged in (Tchekhovskoy et al. 2011;
McKinney et al. 2012). It requires a geometrically thick disk to transport a large mount of flux into the center (Lubow et al. 1994;
Rothstein \& Lovelace 2008; Beckwith et al. 2009; Cao 2011; McKinney et al. 2012). The observed narrow distribution of the jet
production rates of FSRQs implies a universal jet production efficiency among these sources. It remains a problem whether the accretion flow can
accumulate adequate magnetic fields in the inner region to power the jet. More recently, Cao \& Spruit (2013) reported that this problem can be
overcome if the angular momentum of the disk is removed predominantly by the magnetically driven outflows. The thin disk of
a FSRQ may extend to around inner-most stable circular orbit. The so-called magnetically choked accretion flow (MCAF) can form by flux accumulation during a hot, low-accretion-rate phase prior to the cold accretion event (Sikora et al. 2013) and then a powerful jet for FSRQs is produced via the BZ mechanism.

The difference of the jet production mechanisms may also manifest in the observed luminosity, which has been proposed to unify the subclasses of blazars as blazar sequence (Fossati et al. 1998; Ghisellini et al. 1998). It was suggested that the distinction between BL Lacs and FSRQs may be associated with the different accretion rate (Ghisellini et al. 2009a; 2010) since a very weak BLR may form if the accretion rate is lower than  $10^{-2}L_{\rm Edd}$ (e.g., Ho 2008). The BLR thus is also related to the accretion disk structure and the disk radiative efficiency. The division between BL Lacs and FSRQs may be observationally controlled by the luminosity of the BLR measured in Eddington units (Ghisellini et al. 2011; Sbarrato et al. 2012).

\subsection{Jet Composition and Cavity Kinematics}
\begin{figure*}
\includegraphics[angle=0,scale=0.5]{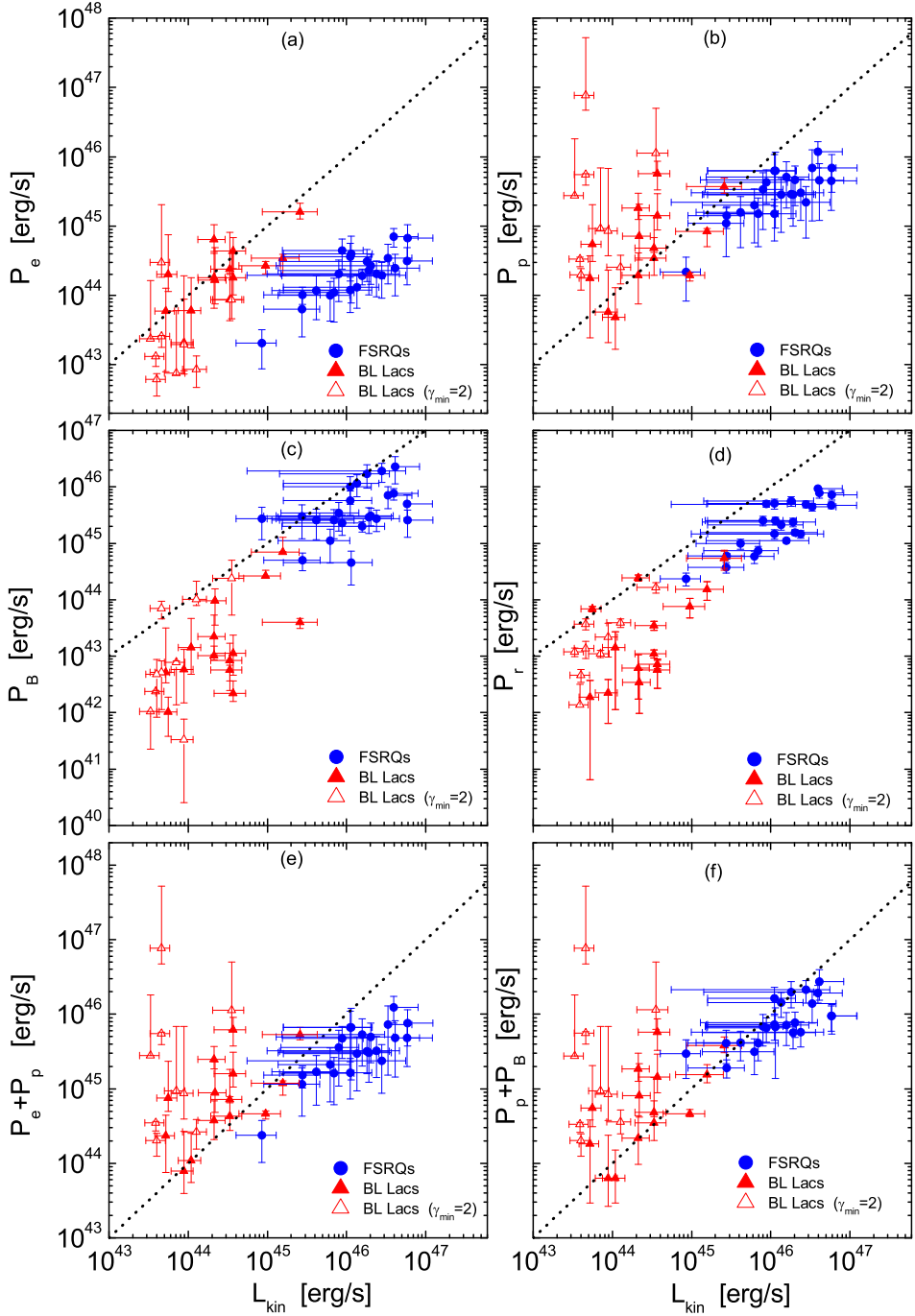}
\caption{Comparisons between $P_{\rm e}$ ({\em panel a}), $P_{\rm p}$ ({\em panel b}), $P_{B}$ ({\em panel c}), $P_{\rm r}$ ({\em panel d}), $P_{\rm e}+P_{\rm p}$ ({\em panel e}), and $P_{B}+P_{\rm p}$ ({\em panel f}) with $L_{\rm kin}$ for FSRQs and BL Lacs. The {\em dotted lines} are the equality lines. }\label{Lkin-Pepbr}
\end{figure*}

It is generally believed that the X-ray cavities are the direct evidence for AGN feedback (B\^{\i}rzan et al. 2008; Cavagnolo et al. 2010),
and the cavity kinetic power is correlated with the radio extended power of galaxies (Rawlings \& Saunders 1991; Willott et al. 1999; B\^{\i}rzan et al. 2004; 2008; Cavagnolo et al. 2010; O'Sullivan et al. 2011;  Meyer et al. 2011). Therefore, in the systems where the X-ray cavities are lacking or not observed, the radio extended luminosity is used to estimate the cavity kinetic power. However, the formation mechanism and composition of the cavity are still unclear. As described in Section 4, we find a strong correlation between $P_{\rm jet}$ and $L_{\rm kin}$ for the blazars in our sample,  indicating that the cavity results from the interaction between AGN jets and the surrounding medium. We further investigate the relation of $L_{\rm kin}$ to $P_{\rm e}$, $P_{\rm p}$, $P_{B}$, $P_{\rm r}$, $P_{\rm e}+P_{\rm p}$, and $P_{B}+P_{\rm p}$ for the FSRQs and BL Lacs in our sample. As shown in Figure \ref{Lkin-Pepbr}, one can observe that $P_{\rm B}$ is comparable with $L_{\rm kin}$ for FSRQs but $P_{\rm p}$ is comparable with $L_{\rm kin}$ for BL Lacs, indicating that the magnetic field energy in the jets may provide the cavity kinetic energy for FSRQs and the kinetic energy of cold protons in the jets may contribute to the cavity kinetic energy of BL Lacs. The kinetic energy of relativistic electrons is not important to produce the cavity for both FSRQs and BL Lacs. This is reasonable since electrons cool off effectively by radiation rather than collisions. Hence, FSRQs and BL Lacs form a well sequence along the equality line in the $P_{B}+P_{\rm p}\sim L_{\rm kin}$ panel of Figure \ref{Lkin-Pepbr}.

The jet composition is also a debated issue. It was suggested that the jets for both FSRQs and BL Lacs are dominated by
particles (e.g., Ghisellini et al. 2009b; 2010). We examine the discrepancy between our results with that of Ghisellini et al. (2010), in which they derived the lepton model parameters for a sample of 53 FSRQs and 31 BL Lacs. As reported in their Table 4, the average values of the parameters are $B=2.6$ G, $\Gamma=13$ (or $\delta=17.8$), $\gamma_{\rm b}=300$, $p_1=1$, and $p_2=2.7$ for FSRQs
and $B=0.8$ G, $\Gamma=15$ ($\delta=18.6$), $\gamma_{\rm b}=1.5E4$, $p_1=1$, and $p_2=3.3$ for BL Lacs. These results are statistically consistent
with ours, i.e., $B=7.4$ G, $\delta=16$, $\gamma_{\rm b}=274$, $p_1=1.6$, and $p_2=3.8$ for FSRQs and $B=0.33$ G, $\delta=18.8$,
$\gamma_{\rm b}=1.5E5$, $p_1=2.0$, and $p_2=3.8$ for BL Lacs. The main difference between our results and theirs is the $\gamma_{\min}$
values. In Ghisellini et al. (2010), $\gamma_{\min}$ is taken as 1 for most of the sources. As shown in Figure \ref{gam_mindis}, the derived $\gamma_{\min}$ for 23 FSRQs ranges from 2 to 86 with a median of 48. Note that the estimates of $P_{\rm p}$ and $P_{\rm e}$
are significantly affected by the $\gamma_{\min}$ values. We use the derived $\gamma_{\min}$ values to calculate the jet powers for the FSRQs in our sample. This significantly lowers both $P_{\rm e}$ and $P_{\rm p}$ values of FSRQ jets in comparison with that reported in Ghisellini et al. (2010). As a result, both radiation efficiency and magnetized parameter of FSRQs derived in our work are systematically higher than those derived in Ghisellini et al. (2010). As described in Section 4, we take $\gamma_{\min}=2$ for those BL Lacs whose $\gamma_{\min}$ cannot be constrained by the SED data. The derived jet powers for most of those BL Lacs tend to be larger than the typical values and significantly
deviate from the fitting line, as also shown in Figures 5, 8, 11, and 12. Therefore, the powers carried by particles
may be overestimated in Ghisellini et al. (2010). Our results show that the jet radiation efficiencies of FSRQs are much higher than that of BL Lacs
and the jets of FSRQs may be dominated by the Poynting flux, whereas the jets of BL Lacs may be dominated by particles.

\begin{figure*}
\includegraphics[angle=0,scale=0.63]{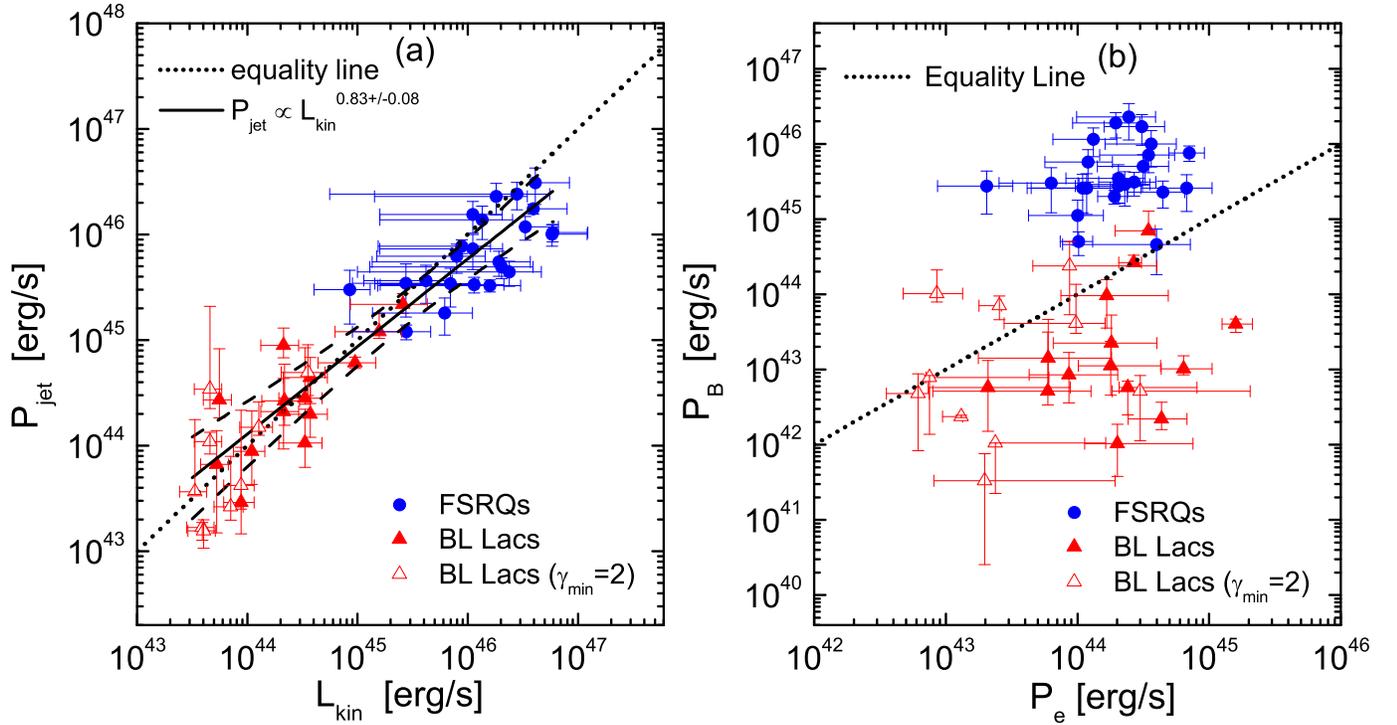}
\caption{ {\em Panel a} --- Comparison between $P_{\rm jet}$ with $L_{\rm kin}$ for FSRQs and BL Lacs, where $P_{\rm jet}$ is assumed to be composed of electron-positron pairs and includes the powers of relativistic electrons ($P_{\rm e}$), magnetic fields ($P_{B}$), and radiation ($P_{\rm r}$), no protons. The {\em solid line} is the best fit for FSRQs and BL Lacs (without considering the opened triangles data points) and the {\em dashed lines} indicate the 3$\sigma$ confidence bands for the best fit. {\em Panel b} --- $P_B$ as a function of $P_{\rm e}$ for FSRQs and BL Lacs in our sample. The {\em dotted lines} in {\em panels a} and {\em b} are the equality lines.}\label{Lkin-PB-Pe}
\end{figure*}

Note that above results are based on the traditional assumption that one cold proton for one relativistic electron in the calculations of jet powers. We check another extreme case that the jet power is carried by positron-electron pairs, magnetic field, and radiation, but no protons. In this scenario, the correlation between jet power $P_{\rm jet}$ and cavity kinetic power $L_{\rm kin}$ becomes tighter with $r=0.93$ and $p=2.2\times10^{-16}$, which is $P_{\rm jet}\propto L_{\rm kin}^{0.83\pm0.08}$, as shown in Figure \ref{Lkin-PB-Pe}(a). Under this extreme scenario, all the FSRQ jets would be dominated by the Poynting flux and most of BL Lac jets are still dominated by the relativistic electrons as shown in Figure \ref{Lkin-PB-Pe}(b).

\section{Summary}
Based on the systematic SED fits with the single-zone leptonic models for the given observed SEDs of GeV-TeV blazars, we have made comparison of the jet properties between the FSRQs and BL Lacs. Our results are summarized below.
\begin{itemize}
\item We derive the $\gamma_{\min}$ value with 1$\sigma$ confidence level for the given observed SEDs of the FSRQs in our sample and find a typically value of $\sim 48$.

\item The magnetic field strengths of the FSRQ jets are larger than that of the BL Lacs by 1-2 orders of magnitude, and the $\gamma_{\rm b}$ values of FSRQs are clustered at several hundreds, being dramatically different from that of BL Lacs, which range in $10^3\sim10^6$, but no statistical difference for the Doppler factors between the FSRQs and BL Lacs is found.

\item Assuming that the total jet power is carried by electron-proton pairs, magnetic field and radiation, we calculate the powers carried by these ingredients and show that the total jet power $P_{\rm jet}$ is correlated with the cavity kinetic power $L_{\rm kin}$ of the jets. Under this assumption, we show that the jet radiation efficiencies of FSRQs are much higher than that of BL Lacs and the jets of FSRQs may be dominated by the Poynting flux, whereas the jets of BL Lacs may be dominated by particles.

\item Different from BL Lacs, a tentative positive correlation between $P_{\rm jet}$ and $M_{\rm BH}$ is found for FSRQs. The jet production and radiation rates per central BH mass for FSRQs, $P_{\rm jet}/L_{\rm Edd}$ and $P_{\rm r}/L_{\rm Edd}$, narrowly range in $0.01\sim0.09$ and $0.002\sim0.023$, respectively. These likely indicate the universal jet production and radiation rates per central BH mass for FSRQs, hence a dominating jet formation mechanism via the BZ process may work at these sources. For the BL Lacs, both $P_{\rm jet}/L_{\rm Edd}$ and $P_{\rm r}/L_{\rm Edd}$ span for several orders of magnitude, may indicate that there is no universal jet production efficiency among these BL Lacs. Therefore there is no a universal jet production mechanism at work among these BL Lacs, i.e., both/either the BP and/or BZ processes, depending on the structures and accretion rates of accretion disks, should operate in BL Lacs.

\item By comparing $L_{\rm kin}$ with $P_{\rm e}$, $P_{\rm p}$, $P_{B}$, $P_{\rm r}$, $P_{\rm e}+P_{\rm p}$, and $P_{B}+P_{\rm p}$ for the FSRQs and BL Lacs in our sample, we find that the magnetic field energy may provide the cavity's kinetic energy for FSRQs, whereas the kinetic energy of cold protons may power the cavities of the BL Lacs.

\end{itemize}

The topic of differences between BL Lacs and FSRQs was extensively covered and discussed (e.g., Ghisellini et al. 2011; Sbarrato et al. 2012). Different from previous papers, we constrain $\gamma_{\min}$ with the observed SEDs and show that the FSRQ jets may be dominated by the Poynting flux and have a high radiation efficiency, whereas the BL Lac jets may be dominated by particles and have a lower radiation efficiency than FSRQs. More essentially, the observed difference of FSRQs and BL Lacs may signal the different jet production mechanisms. The difference of the jet production mechanisms may also manifest in the observed luminosity, as that was proposed to unify the subclasses of blazars as a blazar sequence (Fossati et al. 1998; Ghisellini et al. 1998).

\acknowledgments
We thank the anonymous referee for his/her valuable suggestions. We appreciate helpful discussion with Bing Zhang, G. Ghisellini, Wei Cui, Zi-Gao Dai, Wei-Min Gu, Xin-Wu Cao, Jian-Min Wang, and Yuan Liu. This work is supported by the National Basic Research Program (973 Programme) of China (Grants 2014CB845800), the National Natural Science Foundation of China (Grants 11078008, 11025313, 11363002, 11373036, 11133002), the Strategic Priority Research Program "The Emergence of Cosmological Structures" of the Chinese Academy of Sciences (Grant No. XDB09000000), Guangxi Science Foundation (2013GXNSFFA019001), and Key Laboratory for the Structure and Evolution of Celestial Objects of Chinese Academy of Sciences. Shuang-Nan Zhang acknowledges support from the Qianren start-up grant 292012312D1117210.

\begin{deluxetable}{lcccccccc}
\tabletypesize{\footnotesize}\tablecolumns{16}\tablewidth{40pc} \tablecaption{Data of the BLRs for eight FSRQs in our sample}\tablenum{1}
\tablehead{\colhead{Source}  & \colhead{Flux$_{\rm H\beta}$\tablenotemark{a}}  & \colhead{$\log L_{\rm BLR}$\tablenotemark{a}}
& \colhead{$R_{\rm BLR}$\tablenotemark{b}} &  \colhead{$R_{\rm BLR}$\tablenotemark{c}} &\colhead{$R_{\rm BLR}$\tablenotemark{d}} &
\colhead{$U_{\rm BLR}$\tablenotemark{b}} & \colhead{$U_{\rm BLR}$\tablenotemark{c}} & \colhead{$U_{\rm BLR}$\tablenotemark{d}}
}
\startdata
3C 279&0.8&44.78&83.01&89.61&78.00&3.45&2.96&3.91\\
3C 273&154.8&45.82&329.02&310.72&238.29&2.41&2.70&4.60\\
3C 454.3&4.3&45.94&413.70&382.10&286.93&2.01&2.36&4.18\\
4C 29.45&1.8&45.16&197.80&196.27&157.73&1.46&1.48&2.30\\
PKS 1510-089&5.3&44.87&141.00&144.59&119.86&1.47&1.40&2.04\\
PKS 0208-512&1.2&45.17&245.51&238.55&187.94&0.96&1.02&1.64\\
PKS 0454-234&0.2&44.42&92.74&99.05&85.34&1.20&1.06&1.42\\
PKS 0420-01&0.8&44.92&178.82&179.18&145.33&1.04&1.03&1.57\\
\enddata
\tablenotetext{a}{The data are taken from Celotti et al. (1997), in units of $10^{-14}$ erg s$^{-1}$ cm$^{-2}$. The fluxes of emission line H$\beta$ in PKS 0208-512, PKS 0454-234, and PKS 0420-01 are derived from the fluxes of the emission line $M_{\rm gII}$ using the relative flux in Francis et al. (1991). The luminosities of BLRs for the three sources are also calculated with the luminosities of emission the line $M_{\rm gII}$ [Equation (1) in Celotti et al. 1997].}
\tablenotetext{b}{$R_{\rm BLR}$ (in units of light days) and the corresponding energy densities ($U_{\rm BLR}$, in units of $10^{-2}$ erg cm$^{-3}$) estimated with the relations of $L_{5100}$ to $R_{\rm BLR}$ and $L_{\rm H\beta}$ [Equations (2), (4) in Greene \& Ho 2005].}
\tablenotetext{c}{The same as (b), but using Equation (2) in Wang \& Zhang (2003) with the relative flux between H$\beta$ and H$\alpha$.} \tablenotetext{d}{The same as (b), but using the $R_{\rm BLR}-L_{5100}$ correlation in Bentz et al. (2009) and the $\rm H\beta-L_{5100}$ correlation in Greene \& Ho (2005).}
\end{deluxetable}

\begin{deluxetable}{lcccccccc}
\tabletypesize{\tiny}\tablecolumns{16}\tablewidth{37pc} \tablecaption{The data of the FSRQs in our sample}\tablenum{2} \tablehead{\colhead{Source}  & \colhead{$\log\nu_{\rm s}$}  & \colhead{$\log\nu_{\rm s}f_{\nu_{\rm s}}$}
& \colhead{$\log P_{\rm e}$} &  \colhead{$\log P_{\rm p}$} &\colhead{$\log P_{B}$} & \colhead{$\log P_{\rm r}$} & \colhead{$\log L_{\rm bol}\tablenotemark{\rm a}$} & \colhead{$M_{\rm BH}\tablenotemark{\rm b}$} \\
\colhead{}& \colhead{(Hz)} & \colhead{(erg/cm$^{2}$/s)} & \colhead{(erg/s)}& \colhead{(erg/s)}  & \colhead{(erg/s)} & \colhead{(erg/s)} & \colhead{(erg/s)} & \colhead{$\log M_{\bigodot}$}}
\startdata
3C 279&12.91$\pm$0.15&-10.59$\pm$0.05&44.28$\pm$0.11&45.71$\pm$0.28&45.30$\pm$0.09&45.04$\pm$0.04&47.80$\pm$0.01&9.10$^{\rm C02}$\\
3C 273&13.35$\pm$0.20&-9.55$\pm$0.13&44.00$\pm$0.25&45.30$\pm$0.31&45.05$\pm$0.26&44.77$\pm$0.11&47.11$\pm$0.03&9.30$^{\rm F04}$\\
3C 454.3&12.98$\pm$0.10&-9.93$\pm$0.10&44.85$\pm$0.13&46.07$\pm$0.18&45.88$\pm$0.10&45.97$\pm$0.03&49.06$\pm$0.01&9.64$^{\rm F04}$\\
PKS 1454-354&13.17$\pm$0.40&-10.73$\pm$0.11&44.56$\pm$0.24&45.80$\pm$0.30&46.00$\pm$0.22&45.71$\pm$0.09&48.92$\pm$0.04&\nodata\\
PKS 0208-512&12.86$\pm$0.40&-10.83$\pm$0.13&44.36$\pm$0.24&45.45$\pm$0.28&45.46$\pm$0.21&45.38$\pm$0.08&48.34$\pm$0.02&9.21$^{\rm F04}$\\
PKS 0454-234&12.90$\pm$0.30&-10.76$\pm$0.11&44.12$\pm$0.22&45.45$\pm$0.31&46.06$\pm$0.18&45.33$\pm$0.08&48.53$\pm$0.02&9.17$^{\rm F04}$\\
PKS 0727-11&13.00$\pm$0.20&-11.02$\pm$0.15&44.50$\pm$0.24&45.65$\pm$0.27&45.70$\pm$0.19&45.68$\pm$0.06&48.91$\pm$0.03&\nodata\\
PKS 0528+134&12.80$\pm$0.20&-10.94$\pm$0.14&44.83$\pm$0.24&45.84$\pm$0.24&45.41$\pm$0.22&45.86$\pm$0.09&48.99$\pm$0.06&\nodata\\
4C 66.20 &12.95$\pm$0.48&-10.83$\pm$0.15&44.07$\pm$0.27&45.20$\pm$0.30&45.41$\pm$0.24&44.99$\pm$0.09&47.77$\pm$0.02&9.14$^{\rm W02}$\\
4C 29.45&13.50$\pm$0.25&-10.77$\pm$0.17&44.04$\pm$0.27&45.18$\pm$0.29&45.41$\pm$0.23&44.87$\pm$0.08&47.60$\pm$0.03&9.11$^{\rm C09}$\\
B2 1520+31&13.00$\pm$0.30&-11.40$\pm$0.13&44.31$\pm$0.26&45.53$\pm$0.32&45.54$\pm$0.23&45.41$\pm$0.08&48.64$\pm$0.05&\nodata\\
PKS 0420-01&13.44$\pm$0.30&-10.83$\pm$0.13&44.43$\pm$0.20&45.67$\pm$0.25&45.49$\pm$0.15&45.19$\pm$0.05&48.01$\pm$0.02&9.76$^{\rm C02}$\\
1Jy 1308+326&12.95$\pm$0.35&-11.45$\pm$0.20&44.60$\pm$0.35&45.80$\pm$0.37&44.66$\pm$0.26&45.40$\pm$0.07&48.20$\pm$0.02&8.94$^{\rm C09}$\\
PKS 1510-089&12.95$\pm$0.06&-10.75$\pm$0.05&44.01$\pm$0.12&45.15$\pm$0.15&44.70$\pm$0.15&44.77$\pm$0.05&47.46$\pm$0.03&9.31$^{\rm C02}$\\
4C 28.07&12.91$\pm$0.20&-11.07$\pm$0.17&44.31$\pm$0.23&45.48$\pm$0.26&45.44$\pm$0.18&45.16$\pm$0.07&48.09$\pm$0.02&\nodata\\
PMN 2345-1555&12.87$\pm$0.25&-11.05$\pm$0.13&43.80$\pm$0.26&45.04$\pm$0.29&45.48$\pm$0.26&44.57$\pm$0.09&47.46$\pm$0.04&\nodata\\
S3 2141+17 &13.86$\pm$0.30&-10.06$\pm$0.11&43.31$\pm$0.25&44.34$\pm$0.27&45.44$\pm$0.25&44.37$\pm$0.11&46.78$\pm$0.02&8.98$^{\rm F03}$\\
S4 0133+47&12.95$\pm$0.35&-10.73$\pm$0.13&44.08$\pm$0.23&45.18$\pm$0.26&45.76$\pm$0.20&45.16$\pm$0.08&48.00$\pm$0.01&9.31$^{\rm C02}$\\
S4 0917+44&12.96$\pm$0.30&-11.08$\pm$0.15&44.54$\pm$0.25&45.84$\pm$0.36&45.85$\pm$0.18&45.64$\pm$0.06&48.77$\pm$0.02&9.88$^{\rm C09}$\\
PKS 0227-369&13.10$\pm$0.30&-11.20$\pm$0.13&44.65$\pm$0.20&45.63$\pm$0.22&45.36$\pm$0.17&45.70$\pm$0.06&48.80$\pm$0.03&\nodata\\
PKS 0347-211&13.05$\pm$0.30&-11.11$\pm$0.15&44.29$\pm$0.23&45.34$\pm$0.30&46.28$\pm$0.16&45.69$\pm$0.06&49.13$\pm$0.03&\nodata\\
PKS 2325+093&13.60$\pm$0.30&-10.26$\pm$0.13&44.49$\pm$0.21&45.46$\pm$0.24&46.23$\pm$0.19&45.75$\pm$0.09&48.84$\pm$0.03&\nodata\\
PKS 1502+106&12.95$\pm$0.32&-10.76$\pm$0.14&44.39$\pm$0.26&45.66$\pm$0.32&46.36$\pm$0.22&45.90$\pm$0.08&49.36$\pm$0.04&9.5$^{\rm C09}$\\
\enddata
\tablenotetext{\rm a}{The errors of $L_{\rm bol}$ are estimated with the errors of the two peak luminosities only.}
\tablenotetext{\rm b}{BH masses for the sources. The superscripts denote the references C02: Cao \& Jiang (2002); F04: Fan \& Cao (2004); W02: Woo \& Urry (2002); C09: Chen et al. (2009); and F03: Falomo et al. (2003).}
\end{deluxetable}

\end{document}